\pdfoutput=1
\documentclass[a4paper,12pt]{article}
\usepackage[text={453pt,686pt},centering]{geometry}
\usepackage[noconfig]{refstyle}
\usepackage{lmodern}
\usepackage[T1]{fontenc}
\usepackage{amsmath,amsfonts,amssymb}
\usepackage[sort&compress,square,comma,numbers]{natbib}
\usepackage[usenames,dvipsnames,svgnames,table]{xcolor}
\usepackage{graphicx}
\usepackage[bf]{caption}
\usepackage[colorlinks, citecolor=blue, linkcolor=blue, urlcolor=Maroon, filecolor=Maroon]{hyperref}

\allowdisplaybreaks[4]

\let\eqref=\relax
\newref{eq}{name={eq.~},Name={Eq.~},names={eqs.~},Names={Eqs.~},rngtxt={-},refcmd=(\ref{#1})}
\newref{f}{name={Footnote~},Name={Footnote~},names={Footnotes~},Names={Footnotes~}}
\newref{app}{name={Appendix~},Name={Appendix~},names={Appendixes~},Names={Appendixes~}}
\newref{tab}{name={Table~},Name={Table~},names={Tables~},Names={Tables~}}
\newref{ch}{name={Chapter~},Name={Chapter~},names={Chapters~},Names={Chapters~}}
\newref{sec}{name={Section~},Name={Section~},names={Sections~},Names={Sections~}}
\newref{fig}{name={Figure~},Name={Figure~},names={Figures~},Names={Figures~}}
\numberwithin{equation}{section}

\newcommand{\sfrac}[2]{{\textstyle\frac{#1}{#2}}}
\newcommand{\sfr}[2]{{\textstyle\frac{#1}{#2}}}
\newcommand{\eop}{\mathrm{e}}
\newcommand{\dd}{\mathrm{d}}
\newcommand{\im}{\mathrm{i}}
\DeclareMathOperator\artanh{artanh}
\DeclareMathOperator\diag{diag}
\DeclareMathOperator\vol{vol}

\let\Im=\relax\DeclareMathOperator\Im{Im}
\def\fnote#1#2{\begingroup\def\thefootnote{#1}\footnote{#2}
     \addtocounter{footnote}{-1}\endgroup}

\def\a{\alpha} \def\b{\beta} \def\g{\gamma} \def\d{\delta} \def\e{\epsilon}
  \def\h{\eta} \def\q{\theta}
    \def\m{\mu}
\def\n{\nu}    \def\r{\rho}
 \def\s{\sigma}   \def\f{\varphi}
\def\ff{\phi}  \def\y{\psi} \def\w{\omega}

\def\G{\Gamma} 
\def\D{\Delta} 
   
   \def\L{\Lambda}

\def\F{\Phi}   \def\W{\Omega}

\def\ba{\underline{a}}\def\bb{{\underline{b}}}\def\bc{\underline{c}}
\def\bm{\underline{m}}
\def\bu{\underline{u}}\def\bv{\underline{v}}\def\bw{\underline{w}}

\def\fr{\frac}  \def\dt{\partial}

\def\mc{\mathcal}

\def\Tr{\mbox{Tr}}

\def\Tr{\mbox{Tr}}

\def\RR{\mathbb{R}}

\newcommand\bqa {\begin{eqnarray}}
\newcommand\eqa {\end{eqnarray}}

\newcommand{\bear}{\begin{array}}
\newcommand{\enar}{\end{array}}

\def\beq{\begin{equation}}
\def\eeq{\end{equation}}
\def\bea{\begin{eqnarray}}
\def\eea{\end{eqnarray}}

\begin{document}

\begin{titlepage}

\vfill
\begin{flushright}
\normalsize{ITP--UH--16/14}\\
\normalsize{AEI--2014--041}
\end{flushright}

\vfill

\begin{center}
   \baselineskip=16pt
   {\Large \bf Order $\a'$ heterotic domain walls with \\\vspace{1mm} warped nearly K\"ahler geometry}
   \vskip 2cm
   Alexander~S.~Haupt${}^{1,2}$, Olaf~Lechtenfeld${}^{1}$,
Edvard~T.~Musaev${}^{1,3,4}$
   \vskip .6cm
   \begin{small}
    {\it ${}^{1}$ Institut f\"ur Theoretische Physik and Riemann Center for Geometry and Physics,\\
                  Leibniz Universit\"at Hannover, Appelstra\ss e 2, 30167
Hannover, Germany} \\[0.5cm]
    {\it ${}^{2}$ Max-Planck-Institut f\"ur Gravitationsphysik (Albert-Einstein-Institut),\\
                  Am M\"uhlenberg 1, 14476 Potsdam, Germany} \\[0.5cm]
    {\it ${}^{3}$ Universit\'e de Lyon, Laboratoire de Physique, UMR 5672, CNRS,\\
                  \'Ecole Normale Sup\'erieure de Lyon,\\
                  46 all\'ee d'Italie, 69364 Lyon CEDEX 07, France}\\[0.5cm]
    {\it ${}^{4}$ National Research University Higher School of Economics,
Faculty of Mathematics,\\
20, Myasnitskaya street, 101000 Moscow, Russia}          
   \end{small}
\end{center}

\vfill 
\begin{abstract}
\noindent
We consider $(1{+}3)$-dimensional domain wall solutions of heterotic supergravity on a six-dimensional warped nearly K\"ahler manifold $X_6$ in the presence of gravitational and gauge instantons of tanh-kink type as constructed in~\cite{Harland:2011zs}. We include first order $\a'$ corrections to the heterotic supergravity action, which imply a non-trivial Yang-Mills sector and Bianchi identity. We present a variety of solutions, depending on the choice of instantons, for the special case in which the $SU(3)$ structure on $X_6$ satisfies $W_1^-=0$. The solutions preserve two real supercharges, which corresponds to $\mathcal{N}{=}1/2$ supersymmetry from the four-dimensional point of view. Besides serving as a useful framework for collecting existing solutions, the formulation in terms of dynamic $SU(3)$ structures utilized here allows us to obtain new solutions in as yet unexplored corners of the instanton configuration space. Our approach thus offers a unified description of the embedding of tanh-kink-type instantons into half-BPS solutions of heterotic supergravity where the internal six-dimensional manifold has a warped nearly K\"ahler geometry.
\end{abstract}

\begin{quote}

\end{quote} 
\vfill
\setcounter{footnote}{0}
\setcounter{page}{0}

\fnote{}{alexander.haupt@itp-hannover.de}
\fnote{}{olaf.lechtenfeld@itp-hannover.de}
\fnote{}{emusaev@hse.ru}

\end{titlepage}

\tableofcontents

\section{Introduction}\seclabel{intro}

The problem of finding solutions of ten-dimensional supergravities with a compact manifold filling internal directions is of manifest importance for phenomenological applications (for reviews on that subject, see, for example,~\cite{Grana:2005jc,Wecht:2007wu,Douglas:2006es,Blumenhagen:2006ci,Samtleben:2008pe}). Most of the variety of different vacua in string theory comes from the choice of the internal manifold, the simplest example of which is a flat torus. However, toroidal compactifications lead to lower-dimensional theories that are, in some sense, ``too simple'' in that they typically do not entail realistic phenomenology. In particular, toroidal compactifications of minimally supersymmetric ten-dimensional theories to four dimensions yield $\mc{N}=4$ theories that are non-chiral~\cite{Quevedo:1996sv}. Less trivial examples of manifolds that lead to interesting lower-dimensional physics are given by manifolds with special geometry such as Calabi-Yau (or, more generally, $SU(3)$ structure) manifolds~\cite{Joyce:2001xt}. An important feature of these manifolds is that they preserve less supersymmetry, thus leading to more realistic models. This is a direct consequence of the holonomy principle, which states that the parallel spinor equation
\begin{equation}
 \nabla \e=0
\end{equation}
has $m$ solutions $\e$ if and only if the holonomy group of $\nabla$ is contained in the joint stabilizer subgroup of $m$ spinors, which in turn is related to the $G$ structure of the manifold. Then, $m$ defines the amount of supersymmetry preserved by the background.

In this paper, we study heterotic supergravity, that is the low-energy limit of heterotic string theory, which was first constructed in~\cite{Green:1984sg,Gross:1985fr,Gross:1985rr}. Heterotic supergravity consists of ${\cal N}=1$, $D=10$ supergravity coupled to super Yang-Mills theory. The ingredients are a ten-dimensional manifold ${\cal M}$, equipped with a Lorentzian metric $\hat{g}$, an NS 3-form $\hat{H}$, a dilaton $\hat{\phi}$ and a gauge connection ${}^A \hat{\nabla}$, with gauge group $SO(32)$ or $E_8\times E_8$. It was shown in~\cite{Green:1984sg} that the anomaly cancellation condition of ten-dimensional super Yang-Mills theory coupled to ${\cal N}=1$, $D=10$ supergravity can be written as a Bianchi identity on $\hat{H}$,
\begin{equation}\eqlabel{anomaly}
 \hat{\dd}\hat{H}=\frac{\a'}{4}\Tr(\hat{F}\wedge\hat{F} - \tilde{R}\wedge\tilde{R}) \; ,
\end{equation}
where $\hat{\dd}$ is the ten-dimensional exterior derivative, $\hat{F}$ is the curvature 2-form of the gauge connection ${}^A \hat{\nabla}$ and $\tilde R$ is the curvature 2-form of a connection $\tilde \nabla$. Different connections used in the anomaly cancellation condition~\eqref*{anomaly} correspond to different renormalization schemes~\cite{Hull:1985dx} and there is some discussion about the correct choice of the curvature $\tilde{R}$ in the literature (see, for example,~\cite{Ivanov:2009rh} and references therein). In particular, string theory appears to prefer the choice $\tilde\nabla = {}^+ \hat{\nabla}$~\cite{Bergshoeff:1989de,Becker:2009df}, with ${}^\pm \hat{\nabla}$ being the metric compatible connections on the tangent bundle of ${\cal M}$ with torsion $\pm  \hat{H}$. From a purely supergravity point of view, the connection $\tilde\nabla$ is determined by imposing the instanton equation $\tilde R\cdot \epsilon=0$~\cite{Ivanov:2009rh}. For the purpose of this paper, we will adopt the latter point of view.

In the case of vanishing NS 3-form flux, $\hat{H}=0$, the internal manifold should be Ricci-flat and K\"ahler. Such a solution typically does not stabilize all K\"ahler moduli. Owing to the different scale properties of the terms on the opposite sides of the equation above, a solution with non-zero NS 3-form flux breaks scale invariance and is thus capable of stabilizing the K\"ahler moduli~\cite{Horowitz:1991cd}.  In the present paper, we construct order $\a'$ solutions with non-zero NS 3-form flux that preserve $\mc{N}=1/2$ supersymmetry (that is two real supercharges) in 1+3 external dimensions. The usual $\mc{N}=1$ supersymmetry (that is four real supercharges) implied by the BPS equations is halved by the presence of a domain wall.

At the zeroth order in $\a'$, the BPS equations are solved by $\mathcal{M}=\RR^{1,2}\times c(X_6)$ with vanishing NS 3-form flux $\hat{H}=0$ and $\hat{\ff}=\text{const}$. Here, $c(X_6)$ is the metric cone over a six-dimensional nearly K\"ahler manifold $X_6$. At the first order in $\a'$, the BPS equations can be solved by choosing the gauge connection to be ${}^A \hat{\nabla}={}^{\text{LC}} \nabla$ and $\hat{H}=0$, $\hat{\ff} = \text{const}$~\cite{Harland:2011zs}, where ${}^{\text{LC}} \nabla$ is the Levi-Civita connection on $c(X_6)$. Less trivial solutions with $\hat{H}\neq0$ can be obtained if the gauge field is chosen to be an instanton~\cite{Strominger:1990et,Harvey:1990eg,Khuri:1993ii,Gunaydin:1995ku,Loginov:2008tn,Harland:2011zs,Gemmer:2012pp,Klaput:2012vv}. In our analysis, we exploit the instanton solution of~\cite{Harland:2011zs} and reformulate it in the framework of dynamic $SU(3)$ structures. Moreover, we consider previously unexplored combinations of instanton configurations thereby extending results of~\cite{Harland:2011zs,Gemmer:2012pp,Klaput:2012vv}. In addition, we reproduce some of the solutions found in~\cite{Harland:2011zs,Gemmer:2012pp,Klaput:2012vv}, in special corners of our instanton configuration space. Finally, in the $\a' \to 0$ limit, our construction becomes a sub-sector of the more general zeroth order analysis of~\cite{Gray:2012md}.

The paper is organized as follows. In the next section, we review the $G$ structure formalism for solving the BPS equations of heterotic supergravity, as developed in~\cite{Lukas:2010mf,Gray:2012md}. We also introduce our ansatz for the metric, the NS 3-form and the dilaton and discuss the zeroth order in $\a'$ solution. At the first order in $\a'$, the gauge field $\hat{F}$ couples to the other fields and therefore becomes non-trivial. The subject of \secref{YM} is to review the construction of a certain type of seven-dimensional Yang-Mills instantons that were found in~\cite{Harland:2011zs}. These instantons are employed in order to solve the Yang-Mills sector of the theory. In \secref{warpedNKdomwall}, we combine the seven-dimensional Yang-Mills instantons with the other fields in order to lift the zeroth order solution of \secref{geom} to a fully 10-dimensional solution that is valid at the first order in $\a'$. This involves solving the Bianchi identity and a careful treatment of the equations of motion up to this order. We will also unveil a subtle relationship between static and dynamic $SU(3)$ structures on the six-dimensional compact part $X_6$ of the ten-dimensional space-time manifold, which is a consequence of the warping included in the metric ansatz. In \secref{explsol}, we present explicit solutions assuming that $W_1^-=0$, that is the torsion class $W_1$ of the dynamic $SU(3)$ structure on $X_6$ has vanishing imaginary part. The precise dynamics of the solution depends on the choice of instanton configurations. Our solutions include, in a unified description, special cases of~\cite{Harland:2011zs,Gemmer:2012pp,Klaput:2012vv} and some new ones. We end the main body of the paper by providing a few concluding remarks in \secref{concl}. Finally, our conventions for indices and normalizations are summarized in \appref{conv}.

\section{Geometry of the domain-wall background}\seclabel{geom}

For a background with vanishing fermionic vacuum expectation values to be supersymmetric, the supersymmetry transformations of the corresponding fermionic fields must vanish. This implies certain conditions on the background, known as BPS equations. For the fermionic content of heterotic supergravity, one finds that the BPS equations up to and including terms of order $\a'$ are given by
\begin{equation}\eqlabel{BPS}
\begin{aligned}
 {}^- \hat\nabla \e&=0 \; ,\\
 \left(\hat{\dd}\hat{\ff} - \sfrac{1}{2} \hat{H}\right)\cdot \e&=0 \; ,\\
 \hat{F}\cdot \e&=0 \; ,
\end{aligned}
\end{equation}
for a Majorana-Weyl spinor $\epsilon$. Here and in the following, hatted objects denote ten-dimensional quantities. The conventions used in this paper are summarized in \appref{conv}.

We are interested in the background given by a four-dimensional domain wall with six internal directions filled by a compact manifold $X_6$ with $SU(3)$ structure. We \emph{a priori} choose the following metric ansatz,
\begin{equation}\eqlabel{gen_ansatz}
 \hat{g} = \eop^{2A(x^m)} \left( \h_{\a\b} \dd x^\a \dd x^\b + \eop^{2\D(x^u)} \dd x^3 \dd x^3 + g_{uv}(x^m) \dd x^u \dd x^v \right) \; .
\end{equation}
The world-volume of the domain wall is parametrized by the coordinates $x^\a$ with $\a\in \{0,1,2\}$. The coordinate $x^3$ is chosen to be transverse to the domain wall and will henceforth also be denoted by $y$. The orthonormal frame on the internal six-dimensional manifold $X_6$ is given by $\{e^{\bu}\}=\{e^{\bu}_u \, \dd x^u\}$ with $u\in \{4,5,\ldots,9\}$ and underlined indices denote tangent space (local Lorentz) indices. Finally, the set of coordinates $\{x^m\}=\{x^3,x^u\}$ combines all of the directions transverse to the domain wall world-volume. To summarize, the total ten-dimensional space-time manifold ${\cal M}$ locally splits as
\begin{equation}
 {\cal M}=\RR^{1,2}\times \RR \times X_6 \; ,
\end{equation}
with a flat metric $\h_{\a\b} = \diag(-1,1,1)$ on $\RR^{1,2}$ and a general metric $g_{uv}(x^m)$ compatible with a $y$-dependent $SU(3)$ structure on $X_6$.

As a starting point of our further analysis, we shall briefly repeat here the derivation of $G$ structures consistent with the above ansatz, following~\cite{Lukas:2010mf,Gray:2012md}. First, the Killing spinor $\e$ is decomposed according to our metric ansatz as
\begin{equation}\eqlabel{epsilon}
 \e(x^\a,x^m)=\r(x^\a)\otimes\h(x^m)\otimes \q \; ,
\end{equation}
where $\r$ is the covariantly constant spinor on the world-volume $\RR^{1,2}$ of the domain wall, $\q$ is an eigenvector of the third Pauli matrix and $\h$ is a covariantly constant Majorana spinor on the seven-dimensional space $X_7 := \RR\times X_6$. The spinor $\r$ has two real components, which corresponds to the two real supercharges that our background preserves. In four-dimensional terminology this corresponds to $\mc{N}=1/2$ supersymmetry.

We would like to preserve $(1{+}2)$-dimensional Lorentz invariance on the domain wall world-volume. This restricts $\hat{\ff}$ and $\hat{H}$ such that
\begin{equation}\eqlabel{wvlorinvcond}
 \dt_\a \hat{\ff} =0 \; ,\qquad
 \hat{H}_{\a mn}=0 \; ,\qquad
 \hat{H}_{\a\b n}=0 \; .
\end{equation}
Hence, the only non-zero components of the NS 3-form flux are $\hat{H}_{yuv}$, $\hat{H}_{uvw}$ and $\hat{H}_{\a\b\g}=\ell \e_{\a\b\g}$ with $\e_{\a\b\g}$ being the totally antisymmetric symbol on $\RR^{1,2}$, normalized to $\e_{012}=+1$. Note that this is the same ansatz for $\hat{\ff}$ and $\hat{H}$ as in~\cite{Gray:2012md} and that the system of~\cite{Lukas:2010mf} can be recovered by setting $\hat{H}_{yuv}$ and $\ell$ to zero.

We now proceed by introducing a $G_2$ structure on the seven-dimensional manifold $X_7=\RR\times X_6$ with metric 
\begin{equation}
 g_7 = \eop^{2\D(x^u)} \dd x^3 \dd x^3 + g_{uv}(x^m) \dd x^u \dd x^v \; .
\end{equation}
The $G_2$ structure form $\f \in \L^3 (X_7)$ and its seven-dimensional Hodge dual $\F := \ast_7\f \in \L^4 (X_7)$ are constructed using the seven-dimensional gamma matrices $\Gamma_m$ and the spinor $\h$, which is parallel with respect to ${}^-\nabla$,
\begin{equation}
 \f_{mnp} = - \im \h^\dagger \G_{mnp} \h \; , \qquad \F_{mnpq}=\h^\dagger \G_{mnpq} \h \; .
\end{equation}
The gamma matrices $\G_m$ are taken to satisfy $\{\G_m, \G_n \} = 2 (g_7)_{mn}$, and we define a totally anti-symmetrized product of gamma matrices as $\G_{m_1 \ldots m_p} := \G_{\left[ m_1 \right.} \cdots \G_{\left. m_p \right]}$. The first two equations in~\eqref*{BPS} then imply the following relations~\cite{Lukas:2010mf,Gray:2012md},
\begin{equation}\eqlabel{structureEQ}
\begin{aligned}
 \dd_7 \f &= 2 \dd_7 \hat{\ff} \wedge \f - \ast_7 \hat{H} - \ell \F \; ,\\
 \dd_7 \F &= 2 \dd_7 \hat{\ff} \wedge \F \; ,\\
 \ast_7 \dd_7 \hat{\ff} &= -\sfr12 \hat{H}\wedge \f \; ,\\
 0 &= \sfr12 \ast_7 \ell - \hat{H}\wedge \F \; .
\end{aligned}
\end{equation}
Here, $\dd_7$ and $\ast_7$ are the differential and the Hodge star defined on $X_7$.

Taking into account the decomposition $X_7=\RR\times X_6$, we can rewrite these equations in terms of an $SU(3)$ structure defined on $X_6$ and the domain wall direction. First, we decompose $\eta$ into two six-dimensional spinors of definite chirality,
\begin{equation}
 \h = \sfr{1}{\sqrt{2}} (\h_+ + \h_-) \; .
\end{equation}
An $SU(3)$ structure on $X_6$ is uniquely specified via a real 2-form $J$ and a complex 3-form $\W = \W_+ + \im\W_-$, which are defined for every fixed value of $y$ using the spinors $\eta_\pm$~\cite{Lukas:2010mf,Gray:2012md},
\begin{equation}
 \W_{uvw}=\h_+^\dagger \g_{uvw}\h_- \; ,\qquad
 J_{uv}=\mp \h_{\pm}^\dagger \g_{uv}\h_{\pm} \; ,
\end{equation}
where the $\g_u$ are gamma matrices on $X_6$ satisfying $\{\g_u, \g_v \} = 2 g_{uv}$. The forms $(J,\W)$ obey the following relations,
\begin{equation}\eqlabel{SU3structprops}
 J\wedge\W = 0 \; , \qquad
 \sfrac{1}{3!} J\wedge J\wedge J = \sfrac{\im}{8} \W\wedge\bar\W = \ast 1 \; , \qquad
 \ast J = \sfrac{1}{2} J\wedge J \; , \qquad
 \ast \W_\pm = \pm \W_\mp \; ,
\end{equation}
where $\ast$ is the six-dimensional Hodge star with respect to the metric $g_{uv}(x^m)$. The Hodge star $\ast$ is in our conventions related to $\ast_7$ via
\begin{equation}
 \ast_7 \omega^{(6)}_p = \eop^\D (\ast \omega^{(6)}_p) \wedge \dd y \; , \qquad
 \ast_7 (\dd y \wedge \omega^{(6)}_p) = \eop^{-\D} \ast \omega^{(6)}_p \; .
\end{equation}
Here, $\omega^{(6)}_p$ is a $p$-form with legs only in the $X_6$ directions.

The relation between the $G_2$ structure $(\f,\F)$ and the $SU(3)$ structure $(J,\W)$ can be expressed as~\cite{Lukas:2010mf,Gray:2012md,Chiossi:2002tw}
\begin{equation}\eqlabel{SU3}
\begin{aligned}
\f&=\eop^\D \dd y \wedge J+\W_- \; ,\\
\F&=\eop^\D \dd y\wedge \W_+ + \sfr12 J\wedge J \; ,
\end{aligned}
\end{equation}
where the prefactor $\eop^\D$ is a consequence of the metric ansatz~\eqref*{gen_ansatz}.

Substituting the decomposition~\eqref*{SU3} into~\eqref*{structureEQ}, one obtains
\begin{equation}\eqlabel{structureEQ1}
\begin{aligned}
 \dd J & = \eop^{-\D}\W'_--2\,\eop^{-\D}\hat{\ff}'\W_-+2\,\dd\hat{\ff}\wedge J - J\wedge\dd\D-\ast H + \ell\,\W_+ \; ,\\
 J\wedge \dd J & = J\wedge J \wedge \dd\hat{\ff} \; ,\\
 \dd\W_+ & = \eop^{-\D}J\wedge J' - \eop^{-\D} \hat{\ff}' J\wedge J+2\,\dd\hat{\ff}\wedge \W_+ + \W_+\wedge\dd\D \; ,\\
 \dd\W_-& = 2\,\dd\hat{\ff}\wedge \W_- - \eop^{-\D}\ast H_y -\sfr12 \ell J\wedge J \; ,\\ 
 \ast \dd\hat{\ff} &= \sfr12 \eop^{-\D} H_y \wedge \W_- -\sfr12 H\wedge J \; ,\\
 \ast\hat{\ff}' & = -\sfr12 \eop^{\D} H\wedge \W_- \; ,\\
 0 & = \sfr12 \ast \ell - \W_+\wedge H - \sfr12\eop^{-\D}H_y\wedge J\wedge J \; .
\end{aligned}
\end{equation}
Here, a prime denotes the derivative with respect to the coordinate $y$, and $\dd$ is the exterior derivative on the six-dimensional manifold $X_6$. The two exterior derivatives $\dd_7$ and $\dd$ are related via 
\begin{equation}
 \dd_7 \w = \dd\w + \dd y \wedge \w'
\end{equation}
for some $p$-form $\w$. The ten-dimensional NS 3-form $\hat{H}$ is taken to decompose into the following parts, respecting~\eqref*{wvlorinvcond},
\begin{equation}
 \hat{H} = H + \dd y \wedge H_y + \ell \vol_{\RR^{1,2}} \; ,
\end{equation}
with 
\begin{equation}
 H = \sfr{1}{3!} H_{uvw} \dd x^u \wedge \dd x^v \wedge \dd x^w \qquad\text{and}\qquad
 H_y = \sfr{1}{2!} H_{yuv} \dd x^u \wedge \dd x^v
\end{equation}
having legs solely in the $X_6$ directions. We also define the volume form of $\RR^{1,2}$ as $\vol_{\RR^{1,2}} := \frac{1}{3!} \e_{\a\b\g} \dd x^\a \wedge \dd x^\b \wedge \dd x^\g$.

Note that~\eqref*{structureEQ1} may be regarded as a generalization of the Hitchin flow equations~\cite{Hitchin:2001rw}. This a common situation in four-dimensional BPS domain-wall solutions of ten-dimensional supergravity theories~\cite{Mayer:2004sd,Louis:2006wq,Smyth:2009fu,Lukas:2010mf,Gray:2012md}. In the absence of $\hat{H}$, $\hat\ff$ and $\D$, the system of equations in~\eqref*{structureEQ1} reduces to
\begin{equation}
\begin{aligned}
 J\wedge \dd J &= 0 \; , &\qquad\qquad \dd J &= \W_-' \; , \\
 \dd\W_- &= 0 \; , &\qquad\qquad \dd\W_+ &= J\wedge J' \; ,
\end{aligned}
\end{equation}
which are the original Hitchin flow equations.

The structure forms $J$ and $\W$ are tightly related to the torsion classes defined as irreducible representations of the torsion $T_{mn}{}^p$ under the stability group $SU(3)$. A manifold with $SU(3)$ structure in general has a connection with torsion
\begin{equation}
 T_{mn}{}^p\in \L^1 \otimes \mathfrak{so}(6) \; .
\end{equation}
The 1-form index is the upper index of the torsion tensor, while the lower antisymmetric pair of indices $[mn]$ label an element of $\mathfrak{so}(6) = \mathfrak{su}(3) \oplus \mathfrak{su}(3)^{\perp}$. Decomposing the torsion into irreducible representations of the holonomy group and taking into account that the $\mathfrak{su}(3)$ piece drops out when acting on $SU(3)$-invariant forms, we obtain the intrinsic torsion~\cite{Grana:2005jc}
\begin{equation}
\begin{aligned}
 T^0_{mn}{}^p \in \L^1 \otimes \mathfrak{su}(3)^\perp &= (\bf{3\oplus \bar{3}})\otimes ({\bf 1\oplus 3 \oplus \bar{3}})\\
 & = {\bf (1\oplus 1) \oplus (8\oplus 8) \oplus (\bar{6} \oplus 6) \oplus {\mathrm 2}(3\oplus \bar{3})} \; .\\
 & \hspace{1cm}W_1 \hspace{1.3cm} W_2 \hspace{1.3cm} W_3 \hspace{1cm} W_4,W_5
\end{aligned}
\end{equation}
The tensors $W_1,\ldots,W_5$ are the five torsion classes that appear in the derivatives of the structure forms,
\begin{equation}\eqlabel{dJdW}
\begin{aligned}
 \dd J& = -\sfr32 \Im(W_1 \bar{\W})+W_4 \wedge J + W_3 \; ,\\
 \dd\W & = W_1 J\wedge J + W_2 \wedge J + \bar{W}_5\wedge \W \; .
\end{aligned}
\end{equation}

\vspace{0.1cm}

We now depart from the general discussion and focus on nearly K\"ahler manifolds, which support instanton connections of the type found in~\cite{Harland:2011zs} and are defined by the following condition on the torsion classes,
\begin{equation}
 W_2=W_3=W_4=W_5=0 \; , \qquad\text{while}\qquad W_1 = W_1^+ + \im W_1^-
\end{equation}
is the only non-zero contribution to the intrinsic torsion. In addition, we set 
\begin{equation}
 A=0 \qquad\text{and}\qquad \D=0 \; .
\end{equation}
The system of equations in~\eqref*{structureEQ1} is then solved by an NS 3-form flux and a dilaton of the form~\cite{Gray:2012md},
\begin{equation}\eqlabel{H_full}
\begin{aligned}
 \hat{H} &= -\sfr12 \ff' \W_+ + \left( \sfr32 W_1^- + \sfr78 \ell \right) \W_- - \left(2 W_1^- + \ell \right) J\wedge \dd y + \ell \vol_{\RR^{1,2}} \; , \\
 \hat{\ff} &= \ff(y) \; ,
\end{aligned}
\end{equation}
provided the structure forms $J$ and $\W$ satisfy the following flow and structure equations,
\begin{equation}\eqlabel{structure}
\begin{aligned}
 & J' = (W_1^+ + \ff') J \; , && \dd J = -\sfr32 W_1^- \W_+ + \sfr32 W_1^+ \W_- \; ,\\
 & \W_-' = -\left(3W_1^- + \sfrac{15}{8} \ell \right) \W_+ + \sfr32(W_1^+ + \ff')\W_- \; , && \dd\W = W_1 J\wedge J \; .
\end{aligned}
\end{equation}
By acting with a $y$-derivative on the second equation in~\eqref*{SU3structprops}, one also learns that
\begin{equation}\eqlabel{structure1}
 \W_+' = \sfr32(W_1^+ + \ff')\W_+ + \a(y) \W_- \; ,
\end{equation}
with some as yet undetermined function $\a(y)$. The expressions~\eqrangeref*{H_full}{structure1} represent the most general solution of the first two BPS equations in~\eqref*{BPS} with the general metric ansatz~\eqref*{gen_ansatz} and $A=\D=0$ on a nearly K\"ahler manifold $X_6$. 

Before constructing order $\a'$ solutions, we shall first discuss the zeroth order case. We have already solved the first two BPS equations in~\eqref*{BPS}. In addition, the third BPS equation in~\eqref*{BPS} is solved by $\hat{F}=0$. In order to have a full heterotic supergravity solution, we also need to check that the Bianchi identity and the time-like components of the equations of motion are satisfied. The latter leads to the condition $\ell=0$, as will be shown in more detail in \secref{EOM}. The Bianchi identity at the zeroth order in $\a'$ simply becomes $\hat{\dd}\hat{H}=0$. From this condition, we obtain the following set of equations,
\begin{align}
 0 &= \ff' W_1^+ - 3 (W_1^-)^2 \; , \eqlabel{zerothordereq1} \\
 0 &= \ff'' + \sfr32 (\ff')^2 + \sfrac{13}{2} \ff' W_1^+ \; , \eqlabel{zerothordereq2} \\
 0 &= \ff' \a - 3 (W_1^-)' - \sfrac{21}{2} W_1^+ W_1^- - \sfr92 \ff' W_1^- \; . \eqlabel{zerothordereq3}
\end{align}
We can immediately read off two special solutions
\begin{align}
 &\text{1.} \quad \ff=\text{const.} \; , &W_1^+ &= \text{any} \; , &W_1^- &= 0 \; , &\a &= \text{any}  \; , \\
 &\text{2.} \quad \ff = \sfr23 \log\left(a y + b \right) \; , &W_1^+ &= 0 \; , &W_1^- &= 0 \; , &\a &= 0 \; ,
\end{align}
where $a$, $b$ are integration constants and `any' means a free function. The first case corresponds to a nearly K\"ahler geometry with constant dilaton and vanishing NS 3-form flux. The second case is Calabi-Yau with flux. Both solutions are contained in~\cite{Gray:2012md} as special cases. This concludes our analysis of the zeroth order case, and we shall turn to the construction of order $\a'$ solutions.

\section{Yang-Mills sector}\seclabel{YM}

\subsection{Yang-Mills instantons on \texorpdfstring{$\RR\times X_6$}{R times X6}}\seclabel{inst}

In this section, we review the construction of Yang-Mills instantons \`a la Harland and N\"olle~\cite{Harland:2011zs}. In their terminology, an instanton is a solution of $\hat{F}\cdot\e = 0$, which is the third BPS equation in~\eqref*{BPS}. At the zeroth order in $\a'$, one may simply set $\hat{F}=0$ and ignore the Yang-Mills sector altogether. This is consistent, since the coupling of $\hat{F}$ to the other supergravity fields only starts to arise at linear order in $\a'$. Since our goal is to construct order $\a'$ solutions, we need a non-trivial $\hat{F}$.

We will study the instanton equation on the manifold $X_7=\RR\times X_6$ with `$h$-cone' metric\footnote{The metric~\eqref*{metr_RX6} may be regarded as a generalized cone metric. It reduces to the standard cone metric on $\RR_+\times X_6$ upon setting $h(y)=y$.}
\begin{equation}\eqlabel{metr_RX6}
 g_7 = \dd y^2 + (h(y))^2\,\tilde{g} \; ,
\end{equation}
where $\tilde{g}$ is a fixed (that is $y$-\emph{in}dependent) nearly K\"ahler metric on $X_6$ with components given by $\tilde{g}_{uv}(x^u)$, and $h(y)$ is a warp factor. Note that this metric is further restricted as compared to the $g_7$ introduced in the previous section. We henceforth take $g_7$ to have the form~\eqref*{metr_RX6}.

The orthonormal frame on $X_7$ is given by $\{\s^{\bm}\}=\{\dd y,h\,e^{\bu}\}$ with $\bm=\underline{3},\underline{4},\ldots,\underline{9}$ and $\bu=\underline{4},\underline{5},\ldots,\underline{9}$. Here, $\{e^{\bu}\} = \{e^{\bu}_u \, \dd x^u\}$ is an orthonormal frame on $X_6$ satisfying $e^{\bu}_u e^{\bv}_v \delta_{\bu\bv} = g_{uv}$. Associated to the $y$-independent metric $\tilde{g}$, there is a static (that is $y$-\emph{in}dependent) $SU(3)$ structure on $X_6$ defined in terms of a real 2-form $\tilde{J}$ and a complex 3-form $\tilde{\W}$. The orthonormal frame $\{e^{\bu}\}$ on $X_6$ can be arranged such that $\tilde{J}$ and $\tilde{\W}$ take the following standard form,
\begin{equation}\eqlabel{P}
\begin{aligned}
 \tilde{J}  &= e^{\underline{4}} \wedge e^{\underline{5}} + e^{\underline{6}} \wedge e^{\underline{7}} + e^{\underline{8}} \wedge e^{\underline{9}} \; , \\
 \tilde{\W} &= (e^{\underline{4}} + \im e^{\underline{5}}) \wedge (e^{\underline{6}} + \im e^{\underline{7}}) \wedge (e^{\underline{8}} + \im e^{\underline{9}}) \; .
\end{aligned}
\end{equation}
These forms satisfy $\dd\tilde{\W}_+ = 2 \tilde{J}\wedge\tilde{J}$ and $\dd\tilde{J}=3\tilde{\W}_-$, showing that the manifold $X_6$ is indeed nearly K\"ahler with $\tilde{W}_1^+ = 2$ and $\tilde{W}_1^- = 0$. In addition, they obey the equations in~\eqref*{SU3structprops} with tildes everywhere.

The instanton equation $\hat{F}\cdot\e = 0$ on $X_7$ reduces to
\begin{equation}\eqlabel{inst_diff}
 \ast_7 \hat{F} = - (\ast_7 Q)\wedge\hat{F} \qquad\text{where}\qquad
 Q= h^3 \, \dd y\wedge\tilde{\W}_+ + \sfr12 h^4 \tilde{J}\wedge\tilde{J} \; .
\end{equation}
Via the coordinate redefinition, assuming $h\geq 0$ (see \fref{negative_h}),
\begin{equation}\eqlabel{tau_y_trafo}
 \dd y = \eop^{f(\tau)} \dd\tau \qquad\text{where}\qquad \eop^{f(\tau)} = h(y(\tau)) \; ,
\end{equation}
the metric~\eqref*{metr_RX6} transforms into
\begin{equation}
 g_7 = \eop^{2f} \, g_Z  \qquad\text{with}\qquad g_Z = \dd\tau^2 + \tilde{g} \; ,
\end{equation}
where $g_Z$ is the metric on the cylinder. Since~\eqref*{inst_diff} is conformally invariant and the metric~\eqref*{metr_RX6} is conformal to the cylinder metric, instantons on the cylinder will also solve~\eqref*{inst_diff}. The instanton equation on the cylinder is
\begin{equation}\eqlabel{inst_cylinder}
 \ast_Z \hat{F} = - (\ast_Z Q_Z)\wedge\hat{F} \qquad\text{where}\qquad
 Q_Z = \dd\tau\wedge\tilde{\W}_+ + \sfr12 \tilde{J}\wedge\tilde{J}\; ,
\end{equation}
and $\ast_Z$ is the Hodge star with respect to the cylinder metric $g_Z$.

To solve~\eqref*{inst_cylinder}, we use the same ansatz for the gauge connection as in~\cite{Harland:2011zs}, namely
\begin{equation}\eqlabel{A}
 {}^A \hat\nabla={}^{\text{can}} \nabla + \y(\tau)e^{\bu} I_{\bu} \; ,
\end{equation}
where ${}^{\text{can}} \nabla$ is the canonical connection on $X_6$ defined by
\begin{equation}
 {}^{\text{can}} \w_{u}{}^{\bv}{}_{\bw}={}^{\text{LC}}\w_{u}{}^{\bv}{}_{\bw}+\sfr12 (\tilde{\W}_+)_{\bw\bv\bu} e^{\bu}_u\; ,
\end{equation}
and the matrices $I_{\bu}$, generating the orthogonal complement of $\mathfrak{su}(3)$ in $\mathfrak{g}_2 \subset \mathfrak{so}(7)$ are defined by the following relations,
\begin{equation}\eqlabel{I}
 (I_{\bu})^{\underline{3}}{}_{\bv}=-(I_{\bu})^{\bv}{}_{\underline{3}}=-\d^{\bu}_{\bv} \; ,\qquad
 (I_{\bu})^{\bw}{}_{\bv}=-\sfr12 (\tilde{\W}_+)_{\bu\bv\bw} \; .
\end{equation}
Together with the generators of $\mathfrak{su}(3)$, denoted $(I_{\underline{i}})^{\bu} {}_{\bv}$, they form a basis of the Lie algebra $\mathfrak{g}_2$. The curvature 2-form $\hat{F} = \fr12 [{}^A \hat\nabla, {}^A \hat\nabla ]$ of the gauge connection~\eqref*{A} becomes~\cite{Harland:2011zs}
\begin{equation}\eqlabel{F}
 \hat{F} = {}^{\text{can}} R+\sfr12 \y^2 f^{\underline{i}}_{\bu\bv} e^{\bu} \wedge e^{\bv} I_{\underline{i}} + \dot{\y} \dd\tau\wedge e^{\bu}I_{\bu}+\sfr12(\y-\y^2)(\tilde{\W}_+)_{\bu\bv\bw} e^{\bv} \wedge e^{\bw}I_{\bu} 
 =: \mc{F}(\y) \; ,
\end{equation}
where $f^{\underline{i}}_{\bu\bv}$ is a structure constant appearing in the Lie algebra commutator 
\begin{equation}
[I_{\bu},I_{\bv}]=f^{\underline{i}}_{\bu\bv} I_{\underline{i}} + f^{\bw}_{\bu\bv} I_{\bw}\; ,
\end{equation}
and a dot denotes a derivative with respect to $\tau$. It was shown in~\cite{Harland:2011zs} that such an $\hat{F}$ solves the instanton equation~\eqref*{inst_cylinder} if and only if the function $\y(y)$ satisfies the `kink equation'
\begin{equation}\eqlabel{inst}
 \dot{\y} = 2\,\y\,(\y-1) \; .
\end{equation}
This equation has two fixed points, $\psi=0$ and $\psi=1$, which correspond to the canonical connection ${}^{\text{can}} \nabla$ and the Levi-Civita connection ${}^{\text{LC}} \nabla$ on $X_7$, respectively. There is also a non-constant solution. It interpolates between the two fixed points and is given by the kink function
\begin{equation}\eqlabel{inst_sol}
 \y ( \tau ) = \sfr12 \left( 1 - \tanh[\tau{-}\tau_0] \right) \; .
\end{equation}
The integration constant $\tau_0$ fixes the position of the instanton in the $\tau$ direction. In terms of the original variables $y$ and $h(y)$, the kink equation becomes
\begin{equation}\eqlabel{insty}
 h(y) \, \y'(y) = 2\,\y(y)\,(\y(y)-1) \; .
\end{equation}
It has the same fixed points, $\psi=0$ and $\psi=1$, as~\eqref*{inst}. The non-constant solution is formally given by 
\begin{equation}
 \y(y) = \sfr12 \left( 1 - \tanh[\tau(y){-}\tau_0] \right) \; ,
\end{equation}
with $\tau(y)$ determined by~\eqref*{tau_y_trafo}.

Heterotic supergravity contains another curvature 2-form besides $\hat{F}$, namely $\tilde{R}$. As explained in \secref{intro}, we adopt the supergravity point of view for the purpose of this paper, which implies the instanton equation also for $\tilde{R}$, that is
\begin{equation}
 \tilde{R}\cdot \e = 0 \; .
\end{equation}
Given that we have found an explicit instanton solution for the ansatz~\eqref*{A}, we will make the same ansatz also for $\tilde{\nabla}$. Each connection, however, comes equipped with its own independent scalar function $\y$. To distinguish between the two, we put
\begin{equation}
\tilde{R}=\mc{F}(\y_1) \qquad\text{and}\qquad \hat{F}=\mc{F}(\y_2) \; ,
\end{equation}
where $\mc{F}(\y)$ was defined in~\eqref*{F}. The precise choices for $\y_{1,2}$ will be made later.

\subsection{Bianchi identity}\seclabel{BI}

The Green-Schwarz anomaly cancellation condition for the heterotic string at order $\a'$ can be written as a Bianchi identity for the NS 3-form $\hat{H}$, namely~\cite{Green:1984sg, Ivanov:2009rh}
\begin{equation}\eqlabel{BI}
 \hat{\dd}\hat{H}=\frac{\a'}{4}\Tr(\hat{F}\wedge\hat{F} - \tilde{R}\wedge\tilde{R}) \; .
\end{equation}
With the results from the previous subsection, one may explicitly evaluate the right-hand side of this equation. First, one has
\begin{equation}
 \hat{\dd}\hat{H}=-\frac{\a'}{4}\Tr(\mc{F}(\y_1)\wedge\mc{F}(\y_1) - \mc{F}(\y_2)\wedge\mc{F}(\y_2)) \; .
\end{equation}
Inserting~\eqref*{F} and using the kink equation~\eqref*{insty}, we can express the Bianchi identity in the following form,
\begin{equation}\eqlabel{Bianchi}
 \hat{\dd}\hat{H} = - \fr{\a'}{4}\left[3 h \left( (\y_1')^2 - (\y_2')^2\right)\dd y\wedge\tilde{\W}_+ 
                  + 2 \left( \y_1^2 (2 \y_1 - 3)- \y_2^2 (2 \y_2 - 3) \right)\tilde{J}\wedge\tilde{J}\right] \; ,
\end{equation}
which will turn out to be useful in the following sections.

\section{Warped nearly K\"ahler domain wall}\seclabel{warpedNKdomwall}

\subsection{Static versus dynamic \texorpdfstring{$SU(3)$}{SU(3)} structures on \texorpdfstring{$X_6$}{X6}}

In \secref{geom}, we began our general discussion by assuming a dynamic, that is $y$-dependent, $SU(3)$ structure on $(X_6, g)$ characterized by a pair of forms $(J,\W)$. This was subsequently specialized in \secref{YM} to the case where $g = (h(y))^2\, \tilde{g}$ with a static, that is $y$-independent, $SU(3)$ structure on $(X_6, \tilde{g})$ with the forms $(\tilde{J},\tilde{\W})$.

Given that a pair of $SU(3)$ structure forms uniquely specifies a metric and the relation 
\begin{equation}\eqlabel{gtildeg}
 g = (h(y))^2\,\tilde{g} \; ,
\end{equation}
it is clear that the two pairs $(J,\W)$ and $(\tilde{J},\tilde{\W})$ are not unrelated. Indeed, the Hodge star satisfies $\ast \omega_p = h^{6-2p} \tilde{\ast} \omega_p$ for a $p$-form $\w_p$ on $X_6$, and, together with $\frac{1}{3!} J^3 = \ast 1$ as well as $\frac{1}{3!} \tilde{J}^3 = \tilde{\ast} 1$, this implies
\begin{equation}
 J = h^2 \tilde{J} \; .
\end{equation}
The relation between $\W$ and $\tilde{\W}$ is a little more subtle, due to the fact that there can also be a mixing between real and imaginary parts. To parametrize this mixing, we write
\begin{equation}\eqlabel{OmegaTildeOmegaMixing}
 \W_+ = h^3 \cos\b\,\tilde{\W}_+ + h^3 \sin\b\,\tilde{\W}_- \; , \qquad
 \W_- = -h^3 \sin\b\,\tilde{\W}_+ + h^3 \cos\b\,\tilde{\W}_- \; ,
\end{equation}
with a $y$-dependent mixing angle $\b\in[0,2\pi)$. A shift of $\beta\to\beta{+}\pi$ can be compensated by a sign flip of $h$, and so we may restrict ourselves to $\b\in[0,\pi)$. The chosen parametrization automatically guarantees that
\begin{equation}
\begin{aligned}
 \sfrac{\im}{8}\tilde{\W}\wedge\bar{\tilde{\W}} &= \tilde{\ast}\, 1 \qquad &\implies&& \qquad \sfrac{\im}{8}\W\wedge\bar\W &= \ast 1 \; , \qquad \text{and} \\
 \tilde{\ast}\, \tilde{\W}_\pm &= \pm \tilde{\W}_\mp \qquad &\implies&& \qquad \ast \W_\pm &= \pm \W_\mp \; ,
\end{aligned}
\end{equation}
which is required in order to be compatible with our $SU(3)$ structure conventions as formulated in~\eqref*{SU3structprops}. Finally, comparing the expressions for $\dd J$ and $\dd\W$ with the tilded versions in \secref{inst}, we learn that
\begin{equation}
 W_1^+ = 2 h^{-1} \cos\b  \qquad\text{and}\qquad W_1^- = - 2 h^{-1} \sin\b \; .
\end{equation}
Thus, mixing in~\eqref*{OmegaTildeOmegaMixing} occurs unless either $W_1^+$ or $W_1^-$ vanish. Setting both $W_1^+$ and $W_1^-$ to zero and thereby reducing to a Calabi-Yau geometry is not possible.\footnote{We are interested in solutions where $(X_6, g)$ is compact. Here and in the following, we thus consider only finite $h$ and exclude the possibility of taking the decompactification limit $h\to\pm\infty$.}

Upon inserting the above expressions into the flow equations in~\eqref*{structure}, it follows that
\begin{align}
 h' &= \cos\b + \sfr12 h \ff' \; , \eqlabel{diffeq1} \\
 0 &= h^2 \left(h\b' + 6 \sin\b - \sfrac{15}{8} \ell h \right) \; . \eqlabel{diffeq2}
\end{align}
Moreover, the as yet unknown coefficient function $\a(y)$ in~\eqref*{structure1} is now fixed as
\begin{equation}
 \a(y) = 3 W_1^- + \sfrac{15}{8} \ell = - 6 h^{-1} \sin\b + \sfrac{15}{8} \ell \; .
\end{equation}
The expression~\eqref*{H_full} for the NS 3-form $\hat{H}$ in terms of $(\tilde{J},\tilde{\W})$ becomes
\begin{multline}\eqlabel{H}
 \hat{H} = h^2 \left[ -\sfr12 h \ff' \cos\b + 3 \sin^2 \b - \sfr78 \ell h \sin\b \right] \tilde{\W}_+
         \\ + h^2 \left[ -\sfr12 h \ff' \sin\b - 3 \sin\b\cos\b + \sfr78 \ell h \cos\b \right] \tilde{\W}_-
         \\ + h \left( 4 \sin\b - \ell h \right) \tilde{J}\wedge \dd y + \ell \vol_{\RR^{1,2}} \; .
\end{multline}

It is instructive to pause here and reflect on what we have achieved so far. Provided the coupled ordinary differential equations~\eqrangeref*{diffeq1}{diffeq2} involving the scalar functions $h$, $\b$, $\ell$ and $\ff$ are satisfied, we have a solution of the first two BPS equations in~\eqref*{BPS} for the metric ansatz
\begin{equation}\eqlabel{hatg}
 \hat{g}= \h_{\a\b} \dd x^\a \dd x^\b + \dd x^3 \dd x^3 + (h(x^3))^2\, \tilde{g}_{uv}(x^w) \dd x^u \dd x^v \; ,
\end{equation}
and the restrictions in~\eqref*{wvlorinvcond}. The third BPS equation in~\eqref*{BPS} is solved by the instanton construction presented in \secref{inst}. However, we still need to ensure that the Bianchi identity is satisfied and check that the time-like components of the equations of motion are obeyed. These issues will be addressed in the next two subsections.

\subsection{Embedding of the instanton in the ten-dimensional solution}

In order to embed the instanton solution of \secref{inst} into a fully ten-dimensional solution, we need to impose the Bianchi identity~\eqref*{BI}. In \secref{BI}, we have already computed the right-hand side of the Bianchi identity. The left-hand side can be further specified by applying a ten-dimensional exterior derivative $\hat{\dd}$ on the expression~\eqref*{H} for $\hat{H}$,
\begin{multline}\eqlabel{dH}
 \hat{\dd}\hat{H} = \left\{ \left[ -\sfr12 h^3 \ff' \sin\b - 3 h^2 \sin\b\cos\b + \sfr78 \ell h^3 \cos\b \right]' - 3 h \left( 4 \sin\b - \ell h \right) \right\} \dd y\wedge\tilde{\W}_-
                \\ + \left[ -\sfr12 h^3 \ff' \cos\b + 3 h^2 \sin^2 \b - \sfr78 \ell h^3 \sin\b \right]' \dd y\wedge\tilde{\W}_+
                \\ + 2 \left[ -\sfr12 h^3 \ff' \cos\b + 3 h^2 \sin^2 \b - \sfr78 \ell h^3 \sin\b \right] \tilde{J}\wedge\tilde{J}
                \\ + \ell' \dd y\wedge\vol_{\RR^{1,2}} \; .
\end{multline}
Comparing with~\eqref*{Bianchi}, we obtain the following additional conditions on the scalar functions $h$, $\b$, $\ell$ and $\ff$, now also coupled to the instanton solutions $\y_{1,2}$ as given in~\eqref*{inst_sol},
\begin{align}
 h^3 \ff' \cos\b - 6 h^2 \sin^2 \b + \sfr74 \ell h^3 \sin\b &= \sfrac{\a'}{2} \left(\y_1^2 (2 \y_1 - 3)- \y_2^2 (2 \y_2 - 3)\right) \; , \eqlabel{bianchi_set1} \\
 \left[ h^3 \ff' \sin\b + 6 h^2 \sin\b\cos\b - \sfr74 \ell h^3 \cos\b \right]' &= -6 h \left( 4 \sin\b - \ell h \right) \; , \eqlabel{bianchi_set2} \\
 \ell' &= 0 \; . \eqlabel{bianchi_set3}
\end{align}
From the $\dd y\wedge\tilde{\W}_+$ term in~\eqref*{dH}, one obtains also the $y$-derivative of~\eqref*{bianchi_set1}. This, however, yields no further condition and is thus omitted. The relation \eqref*{bianchi_set2} may be re-written in the form
\begin{multline}
  \sin\b \left( h^2 \ff'' + \sfr32 (h\ff')^2 + 27 h \ff' \cos\b 
 + \sfrac{105}{32} \ell^2 h^2 - \sfrac{111}{4} \ell h \sin\b \right. \\ \left.
 + 12 \sin^2 \b + 48 \vphantom{\sfr32}\right) - \sfr34 \ell h^2 \ff' \cos\b = 0 \; ,
\end{multline}
after using~\eqref*{diffeq1},~\eqref*{diffeq2} and~\eqref*{bianchi_set3}.

\subsection{Equations of motion}\seclabel{EOM}

The equations of motion of heterotic supergravity up to and including terms of order $\a'$ read
\begin{equation}\eqlabel{eom}
\begin{aligned}
\hat{R}_{\mu\nu} +2 (\hat{\nabla} \hat{\dd}\hat{\phi})_{\mu\nu} 
-\sfrac14 \hat{H}_{\kappa\lambda\mu}\hat{H}_{\nu}{}^{\kappa\lambda}
+\sfrac {\alpha'}4 \left[ \tilde R_{\mu \kappa\lambda\sigma} \tilde R_\nu {}^{\kappa\lambda\sigma} 
- \text{tr}\left(\hat{F}_{\mu\kappa} \hat{F}_\nu {}^\kappa\right)\right]&=0 \; , \\[4pt]
\hat{R} + 4 \hat{\Delta}\hat{\phi}  -4|\hat{\dd}\hat{\phi}|^2 -\sfrac12 |\hat{H}|^2
+\sfrac {\alpha'} 4\text{tr}\left[ |\tilde R|^2 - |\hat{F}|^2 \right]&=0 \; , \\[4pt]
\eop^{2\hat{\phi}}\hat{\dd}\,\hat{\ast}\, (\eop^{-2\hat{\phi}}\hat{F}) + \hat{A}\wedge\hat{\ast}\hat{F} - \hat{\ast}\hat{F}\wedge\hat{A} + \hat{\ast}\hat{H}\wedge\hat{F} &=0 \; , \\[4pt]
\hat{\dd}\,\hat{\ast}\,\eop^{-2\hat{\phi}}\hat{H} &=0 \; .
\end{aligned}
\end{equation}
Here, $\hat{R}_{\mu\nu}$ and $\hat{R}$ are the Ricci tensor and the scalar curvature, respectively. They are computed from the full ten-dimensional metric~\eqref*{hatg}. The gauge field $\hat{A}$ corresponds to the curvature $\hat{F}=\mc{F}(\y_2)$, and its components can be read off from 
\begin{equation}
{}^A \hat\nabla={}^{\text{can}} \nabla + \y_2(y)e^{\bu} I_{\bu} = \hat{\dd} + \hat{A}\; . 
\end{equation}
We will, however, not need the explicit form of this field. Finally, we define $|\w|^2 := \frac{1}{p!} \w_{\mu_1 \ldots \mu_p} \w^{\mu_1 \ldots \mu_p}$ for a $p$-form $\w$, and we note that the Einstein equation has been simplified by means of the dilaton equation.

Since we adopted the supergravity point of view on the curvature $\tilde{R}$, we do not need to verify explicitly all the equations of motion. The precise implication for the equations of motion following from the BPS equations and Bianchi identity is a somewhat subtle point~\cite{Ivanov:2009rh, Gauntlett:2002sc, Martelli:2010jx, Gray:2012md} (for recent discussions, see also~\cite{delaOssa:2014cia, Fernandez:2014kwa, Melnikov:2014ywa, Maxfield:2014wea}). We remark that ans\"atze for the connections and various other assumptions, such as the considered order in the $\a'$~expansion, differ in the literature, and this effects the conclusions that are drawn. For the purpose of this paper, we shall follow a conservative strategy and assume that, for our set-up, the BPS equations together with the Bianchi identity and the \emph{time-like} components of the equations of motion imply the remaining components of the equations of motion.

The time-like components of the Yang-Mills equation are trivially satisfied by the an\-satz, since the time-like components $\hat{F}_{0\m}$ of the corresponding field strength are identically zero. The same is true for the mixed $(0\m)$-components of the Einstein equations where $\m\neq 0$. Hence, we are left with the following two equations,
\begin{align}
 \hat{R}_{00}+2(\hat{\nabla}\hat{d}\hat{\ff})_{00}-\sfr14 \hat{H}_{0\m\n} \hat{H}_0{}^{\m\n}=0 \; , \eqlabel{eom_red1} \\
 \hat{\nabla}_\m \left(\eop^{-2\hat{\ff}} \hat{H}^{\m\n 0}\right)=0 \; . \eqlabel{eom_red2}
\end{align}
For our field ansatz,~\eqref*{eom_red2} is satisfied trivially, whereas~\eqref*{eom_red1} implies
\begin{equation}
 \ell = 0 \; .
\end{equation}
This condition eliminates any flux that has legs in the domain-wall world-volume directions.

\subsection{Summary of the system of equations}

For the warped nearly K\"ahler domain wall considered in this paper, we have obtained in the previous sections a system of coupled non-linear ordinary differential equations for the scalar functions $h$, $\b$ and $\ff$. This system of equations is a consequence of consistently embedding the instanton on $\RR\times X_6$ into a full-fledged heterotic supergravity solution with NS 3-form flux given by~\eqref*{H}. Before we study explicit solutions, it is beneficial to first collect and summarize the complete system of equations. This shall then serve as the central point of reference for the remainder of this paper.

Having set $\ell=0$ in accordance with the result of the previous subsection, the full set of equations then reads as follows,
\begin{align}
 h' &= \cos\b + \sfr12 h \ff' \; , \eqlabel{fulleqsys1} \\
 0 &= h^2 ( h\b' + 6 \sin\b ) \; , \eqlabel{fulleqsys2} \\
 0 &= h^3 \ff' \cos\b - 6 h^2 \sin^2 \b - \sfrac{\a'}{2} \left(\y_1^2 (2 \y_1 - 3)- \y_2^2 (2 \y_2 - 3)\right) \; , \eqlabel{fulleqsys3} \\
 0 &= \sin\b \left( h^2 \ff'' + \sfr32 (h\ff')^2 + 27 h \ff' \cos\b + 12 \sin^2 \b + 48 \right) \; , \eqlabel{fulleqsys4} \\
 h\, \y_{1}' &= 2\,\y_{1}\,(\y_{1}-1) \; , \eqlabel{fulleqsys5} \\
 h\, \y_{2}' &= 2\,\y_{2}\,(\y_{2}-1) \; . \eqlabel{fulleqsys6}
\end{align}
This is an \emph{a priori} over-determined system since there are six equations for the five scalar functions $h$, $\b$, $\ff$, $\y_1$ and $\y_2$. It is supplemented by the determination of the torsion class $W_1$ of the warped six-dimensional space $(X_6, g)$ via $W_1 = 2 h^{-1} \, \eop^{-\im\b}$.

A solution obeying~\eqrangeref*{fulleqsys1}{fulleqsys6} guarantees that the corresponding ten-di\-men\-sio\-nal fields $\hat{g}$, $\hat{H}$, $\hat{\ff}$ and ${}^A \hat{\nabla}$ satisfy the BPS equations~\eqref*{BPS}, the Bianchi identity~\eqref*{BI} and, in turn, also the equations of motion~\eqref*{eom}. Since the connection ${}^- \hat\nabla$ determined by the NS 3-form flux~\eqref*{H} preserves one parallel spinor $\e$ as in~\eqref*{epsilon}, we are dealing with an $\mc{N}{=}1/2$ solution from the point of view of $\RR^{1,3}$ spanned by the non-compact coordinates $\{x^0,x^1,x^2,x^3\}$.

An analytic solution to the highly complicated system of equations~\eqrangeref*{fulleqsys1}{fulleqsys6} would be too much to ask for. Instead, one may concentrate on special cases and hope to find analytic or numerical solutions there. An obvious specialization is to turn off either $W_1^+$ or $W_1^-$.

The first case, that is $W_1^+ = 0$, can be dealt with very quickly. Setting $W_1^+ = 0$ and assuming $h<\infty$ implies $\cos\b=0$. However,~\eqref*{fulleqsys2} then immediately leads to $h = 0$, which causes the metric to be ill-defined. We thus conclude that a well-defined solution requires $W_1^+ \neq 0$.

Fortunately, the second case, that is $W_1^- = 0$, turns out to be more fruitful. It will be the subject of the next section.

\section{Explicit solutions with \texorpdfstring{$W_1^- = 0$}{vanishing imaginary part of the first torsion class}}\seclabel{explsol}

In this section, we investigate the solution of~\eqrangeref*{fulleqsys1}{fulleqsys6} for the special case that $W_1^-=0$. For $h<\infty$ this implies $\sin\b=0$, hence $\cos\b=1$. (Here, we are restricting to $\b\in[0,\pi)$, without loss of generality, as explained below~\eqref*{OmegaTildeOmegaMixing}.) Therefore, the relation~\eqref*{OmegaTildeOmegaMixing} involves no mixing,
\begin{equation}
\W_+ = h^3 \tilde{\W}_+ \qquad\text{and}\qquad \W_- = h^3 \tilde{\W}_-\; .
\end{equation}
The system of equations in~\eqrangeref*{fulleqsys1}{fulleqsys6} then reduces to
\begin{align}
 h' &= 1 + \sfr12 h \ff' \; , \eqlabel{W1p_ODE1} \\
 h^3 \ff' &= \sfrac{\a'}{2} \left(\y_1^2 (2 \y_1 - 3)- \y_2^2 (2 \y_2 - 3)\right) \; , \eqlabel{W1p_ODE2} \\
 h\, \y_{1}' &= 2\,\y_{1}\,(\y_{1}-1) \; , \eqlabel{W1p_ODE3} \\
 h\, \y_{2}' &= 2\,\y_{2}\,(\y_{2}-1) \; , \eqlabel{W1p_ODE4}
\end{align}
and the NS 3-form flux is simply given by
\begin{equation}
 \hat{H} = - \sfr12 h^3 \ff' \tilde{\W}_+ = - \sfrac{\a'}{4} \left(\y_1^2 (2 \y_1 - 3)- \y_2^2 (2 \y_2 - 3)\right) \tilde{\W}_+ \; .
\end{equation}
We remark that there are no terms containing $\tilde{\W}_-$ and $\tilde{J}\wedge \dd y$. Indeed, in the absence of $\ell$, those contributions are precisely measured by $W_1^-$, which is taken to vanish here. 
Hence, the NS 3-form flux is completely internal, that is $H_y=0$. The solution then belongs to the same class of half-flat constructions as those obtained in~\cite{Klaput:2012vv}, given the condition on the dilaton $\dd\ff=0$. The precise forms of the solutions, however, depend on the instanton configurations in question. The relation to solutions contained in the existing literature will be clarified below.

Following~\cite{Harland:2011zs,Gemmer:2012pp}, we may integrate~\eqrangeref*{W1p_ODE1}{W1p_ODE4} after performing a combined coordinate and function redefinition $y\to\tau$, $h(y)\to f(\tau)$ of the form\footnote{\flabel{negative_h}Note that this transformation requires $h$ to be non-negative. However, for the solutions discussed below this is not always the case. For the cases with $\y_1=0,1$ and $\y_2=0,1$, one may solve~\eqrangeref*{W1p_ODE1}{W1p_ODE4} directly and this yields a solution valid for all $y\in\RR$. In cases where at least one of the $\psi_{1,2}$ is a non-constant solution of~\eqrangeref*{W1p_ODE3}{W1p_ODE4} and when expressed in terms of $y$, our solutions are only valid on a half-space ranging from the physical location of the domain wall to infinity. It is then interesting to ask whether and how the solution can be continued across the domain wall. A general argument for the existence of such a continuation for all $y\in\RR$ has been given in~\cite{Gemmer:2012pp}, Section 5.6. We will, however, not attempt to answer this question in the present paper.}
\begin{equation}\eqlabel{y_tau_trafo}
 \dd y = \eop^{f(\tau)} \dd\tau \; , \qquad h(y) = \eop^{f(\tau(y))} \; .
\end{equation}
The metric then becomes
\begin{equation}
 \hat{g}= \h_{\a\b} \dd x^\a \dd x^\b + \eop^{2f(\tau)} \left( \dd\tau^2 + \tilde{g}_{uv}(x^w) \dd x^u \dd x^v \right) \; ,
\end{equation}
and~\eqrangeref*{W1p_ODE1}{W1p_ODE4} turn into
\begin{align}
 \dot\ff &= 2 (\dot{f} - 1) \; , \eqlabel{W1p_ODE1_tau} \\
 2(\dot{f} - 1) \eop^{2(f - \tau)} &= \sfrac{\a'}{2} \eop^{- 2 \tau} \left(\y_1^2 (2 \y_1 - 3)- \y_2^2 (2 \y_2 - 3)\right) \; , \eqlabel{W1p_ODE2_tau} \\
 \dot{\y}_1 &= 2\,\y_{1}\,(\y_{1}-1) \; , \eqlabel{W1p_ODE3_tau} \\
 \dot{\y}_2 &= 2\,\y_{2}\,(\y_{2}-1) \; , \eqlabel{W1p_ODE4_tau}
\end{align}
where a dot denotes a derivative with respect to the new coordinate $\tau$. Equation~\eqref*{W1p_ODE1_tau} implies
\begin{equation}\eqlabel{solphitau}
 \ff (\tau) = \ff_0 + 2 (f - \tau) \; ,
\end{equation}
with a constant of integration denoted $\ff_0$. Equation~\eqref*{W1p_ODE2_tau} can be integrated as well, and we find
\begin{equation}\eqlabel{solf}
 \eop^{2f} = \eop^{2(\tau-\tau_0)} + \sfrac{\a'}{4} \left( \y_1^2 - \y_2^2 \right) \; ,
\end{equation}
with some constant of integration denoted $\tau_0$. The different choices for $\y_{1,2}$ are either $0$, $1$ or the interpolating kink solution~\eqref*{inst_sol} with integration constants $\tau_{1,2}$. This leaves us with a total of 8 different instanton configurations, which are discussed below.

\paragraph{Case 1: $\y_1 = \y_2$.}
All $\a'$ corrections cancel out and we revert to the zeroth order solution of~\cite{Klaput:2012vv}, Section 3.6, that is
\begin{equation}
 f = \tau-\tau_0 \; , \qquad \ff = \ff_0 - 2 \tau_0 \; .
\end{equation}
This is shown in \figref{case1}. The NS 3-form flux vanishes, $\hat{H} = 0$, the dilaton is constant, and the ten-dimensional metric
\begin{equation}
 \hat{g}= \h_{\a\b} \dd x^\a \dd x^\b + \dd x^3 \dd x^3 + (x^3)^2 \tilde{g}_{uv}(x^w) \dd x^u \dd x^v
\end{equation}
is the cone metric on $X_7$ together with the flat Minkowski metric on $\RR^{1,2}$ (ignoring the irrelevant integration constant $\tau_0$, which can always be absorbed by a scale transformation of the coordinate $x^3 := \eop^{\tau-\tau_0}$).
\begin{figure}[t]
\centering
\includegraphics[width=0.7\textwidth]{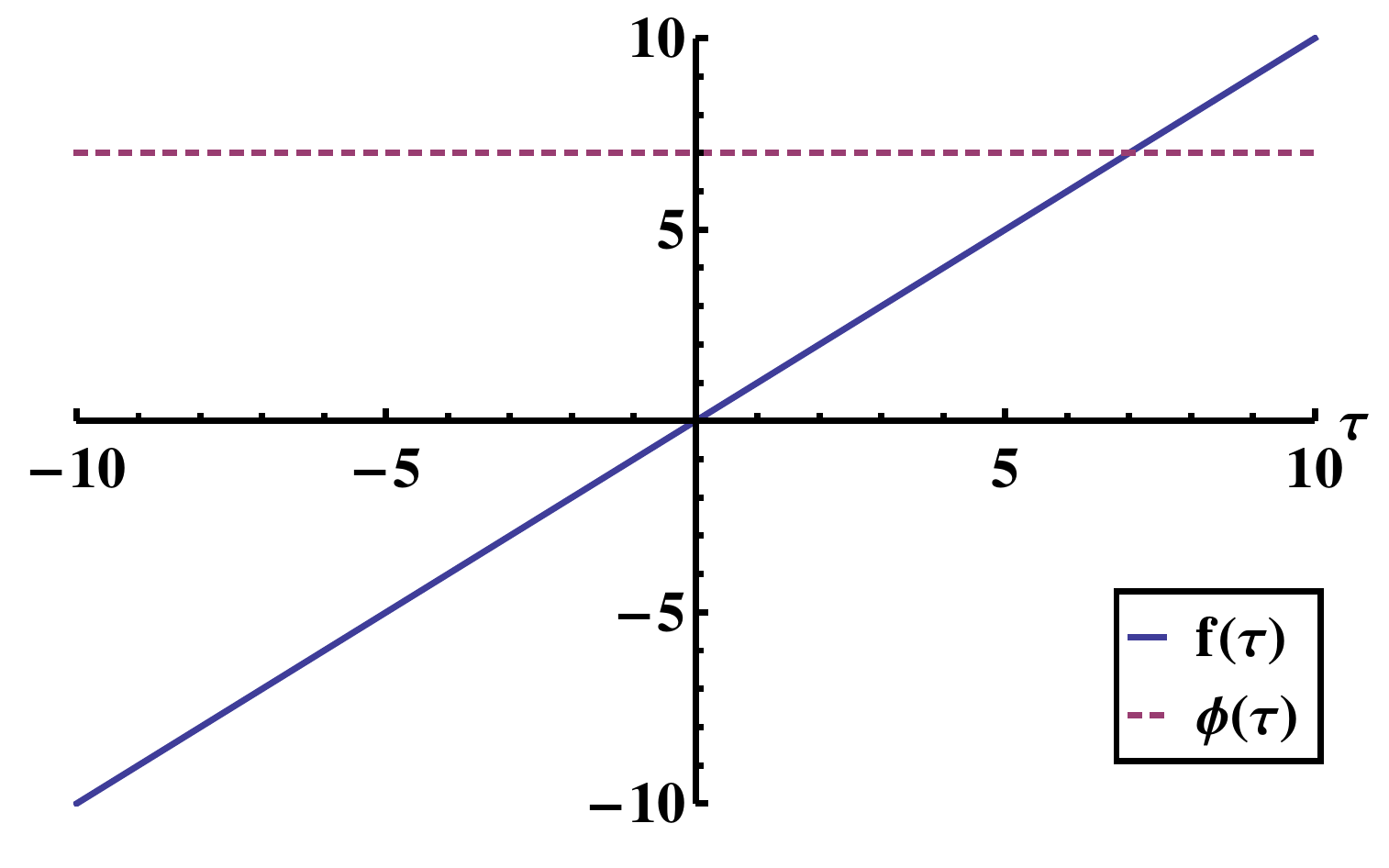}
\caption{Solution for case 1 ($\y_1 = \y_2$). This is the trivial solution corresponding to a cone metric with constant dilaton and vanishing NS 3-form flux.}
\figlabel{case1}
\end{figure}

\paragraph{Case 2: $\y_1 = 1$, $\y_2 = 0$.}
This case admits a static solution with a linear dilaton,
\begin{equation}
f = f_0 := \sfr12 \log(\sfrac{\a'}{4}) \qquad\text{and}\qquad 
\ff = \ff_0 + 2 (f_0{-}\tau)\;.
\end{equation}
The general solution can immediately be stated by inserting the values for $\y_{1,2}$ into~\eqrangeref*{solphitau}{solf},
\begin{equation}\eqlabel{case2_solf}
 \eop^{2f} = \eop^{2(\tau-\tau_0)} + \sfrac{\a'}{4}  \qquad\text{and}\qquad
 \eop^{\ff - \ff_0} = \eop^{- 2\tau_0} + \sfrac{\a'}{4} \eop^{- 2\tau} \; .
\end{equation}
This is a special case of~\cite{Harland:2011zs}, Section 5.1 (with $\psi=0$) and of~\cite{Gemmer:2012pp}, Section 5.5 (with $Q_e = 0$, $a=1$ and taking the decompactification limit in the $S^1$~direction). An exemplary plot showing the behavior of $f(\tau)$ and $\ff(\tau)$ is shown on the left of \figref{case2}.

The transformation back to the original variable $y$ is problematic, however. Solving~\eqref*{y_tau_trafo}, with the solution~\eqref*{case2_solf} inserted, formally yields
\begin{equation}
 y(\tau) = \sfr12 \left(\sqrt{4\,\eop^{2 (\tau - \tau_0)} + \a'} - \sqrt{\a'}\,\artanh\left[\sqrt{1+\sfrac{4}{\a'}\,\eop^{2 (\tau - \tau_0)}}\right] \right) + y_0 \; .
\end{equation}
For finite $\tau$, the argument of $\artanh$ is always greater than one and thus $y(\tau)$ is ill-defined over the reals. One may however directly solve~\eqrangeref*{W1p_ODE1}{W1p_ODE4} for this case. After inserting~\eqref*{W1p_ODE2} into~\eqref*{W1p_ODE1} and considering the inverse function $y(h)$, we obtain
\begin{equation}\eqlabel{case2_invhy}
 y - y_0 = h - \frac{\sqrt{\a'}}{2} \artanh\left(\frac{2h}{\sqrt{\a'}}\right) \; .
\end{equation}
Another inversion, which however cannot be performed explicitly, yields $h(y)$. We see that in terms of the original variables $y$ and $h(y)$, this case is equivalent to the scenario discussed in~\cite{Klaput:2012vv}, Section 4.5.2. 

The graph of $h(y)$ has a kink shape and a zero at the value $y=y_0$, which indicates the location of the domain wall. In addition,~\eqref*{case2_invhy} does not have solutions for all values of $y$. Instead, $y$ is constrained to lie in the interval $|y-y_0| \leq y_{\text{max}}$. The boundary value $y_{\text{max}}$ can only be determined numerically. It depends on $\a'$ and $y_0$. As $y\to\pm y_{\text{max}}$, $h(y)$ limits to $\mp\frac{\sqrt{\a'}}{2}$. We note that the scalar curvature for the metric~\eqref*{hatg} given by the following expression,
\begin{equation}\eqlabel{scalRy}
 \hat{R} = -12 h^{-1} h'' - 30 h^{-2} (h')^2 + h^{-2} \tilde{R} \; ,
\end{equation}
diverges at the location of the domain wall, $y=y_0$, and we thus expect the supergravity approximation to break down in the vicinity of the domain wall. In the expression above, $\tilde{R}$ denotes the scalar curvature of the static nearly K\"ahler metric $\tilde{g}$ on $X_6$, which is conventionally normalized to $\tilde{R}=30$~\cite{Gray:1976}.

The solution for $\ff(y)$ can be implicitly written as
\begin{equation}
 \ff(h) = \log\left( \frac{h^2}{4 h^2 - \a'} \right) + \ff_0 \; ,
\end{equation}
where it is understood that the solution for $h(y)$ should be inserted. Taken at face value, this solution is ill-defined as a function of $y$, even inside the range $|y-y_0| \leq y_{\text{max}}$, since the argument inside the $\log$ is negative. This can be cured by using that $\log(-|x|) = \log |x| + i\pi$, and absorbing the second term into a new integration constant $\tilde{\ff}_0 = \ff_0 + i\pi$. The solution for $\ff$ then reads as follows,
\begin{equation}
 \ff(h) = \log\left( \Big|\frac{h^2}{4 h^2 - \a'} \Big| \right) + \tilde{\ff}_0 \; .
\end{equation}
As $y\to \pm y_{\text{max}}$ we have $\ff \to \infty$, and as $y\to y_0$ we have $\ff \to -\infty$. The plots of $h(y)$ and $\ff(y)$, ignoring the integration constant $\tilde{\ff}_0$, are shown on the right of \figref{case2}.
\begin{figure}[t]
\centering
\includegraphics[width=\textwidth]{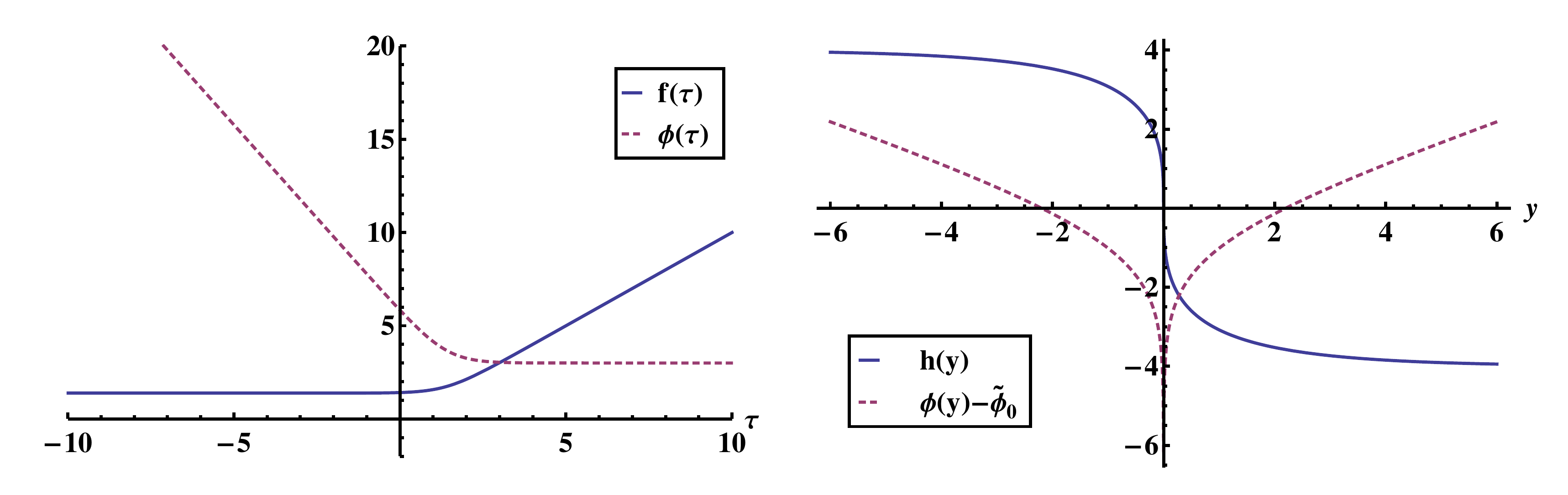}
\caption{Solution for case 2 ($\y_1 = 1$, $\y_2 = 0$). The solution is plotted in terms of the new coordinate $\tau$ on the left and in terms of the old coordinate $y$ on the right. We have adjusted the free constants to $\a'=64$, $\tau_0=0$, $\ff_0 = 3$ and $y_0 = 0$, for concreteness.}
\figlabel{case2}
\end{figure}

\paragraph{Case 3: $\y_1 = 0$, $\y_2 = 1$.}
Contrary to the previous case, there is no static solution in this case. The general solution~\eqrangeref*{solphitau}{solf} now becomes
\begin{equation}\eqlabel{solf_case3}
 \eop^{2f} = \eop^{2(\tau-\tau_0)} - \sfrac{\a'}{4} \; , \qquad
 \eop^{\ff - \ff_0} = \eop^{- 2\tau_0} - \sfrac{\a'}{4} \eop^{- 2\tau} \; .
\end{equation}
This case appeared neither in~\cite{Harland:2011zs} nor in~\cite{Gemmer:2012pp}. It should be noted that the expression for $f(\tau)$ following from~\eqref*{solf_case3} is ill-defined for $\tau < \tau_{\text{dw}}$, where $\tau_{\text{dw}} = \tau_0 + \fr12 \log(\frac{\a'}{4})$ is the location of the domain wall. For $\tau\to \tau_{\text{dw}}$ from above, the scalar curvature $\hat{R}$, given in terms of the new variables $\tau$ and $f(\tau)$ by\footnote{This expression is obtained from~\eqref*{scalRy} by applying the transformation~\eqref*{y_tau_trafo}.}
\begin{equation}\eqlabel{scalRtau}
 \hat{R} = -6\,\eop^{-2f} \left( 2\ddot{f} + 5 \dot{f}^2 \right) + \eop^{-2f} \tilde{R} \; ,
\end{equation}
is divergent, indicating a breakdown of the supergravity approximation in this region. When transformed back to the original variables $y$ and $h(y)$, the solution~\eqref*{solf_case3} is valid only in the half-space $y\in[y_0,\infty)$, where $y_0 = y(\tau_{\text{dw}})$ is the physical location of the domain wall in the $y$ coordinate.

As in the previous case, we obtain a solution for all $y\in\RR$ by directly solving~\eqrangeref*{W1p_ODE1}{W1p_ODE4}. We find
\begin{equation}
 y - y_0 = h - \frac{\sqrt{\a'}}{2} \arctan\left(\frac{2h}{\sqrt{\a'}}\right) \; .
\end{equation}
This case can also be found in~\cite{Klaput:2012vv}, Section 4.5.3. Again, the graph of $h(y)$ has a kink shape and a zero at the value $y=y_0$, which indicates the location of the domain wall. For $y\to\pm\infty$, the function $h(y)$ becomes approximately linear $h(y) \approx y$. The solution for $\ff(y)$ is given implicitly by
\begin{equation}
 \ff(h) = \log\left( \frac{h^2}{4 h^2 + \a'} \right) + \ff_0 \; .
\end{equation}
The graph of $\ff(y)$ has a singularity, $\ff\to-\infty$, at the domain wall location $y=y_0$ and approaches the limiting value $\phi \to \log(\fr14) + \ff_0$ as $y\to\pm\infty$. The plots of $h(y)$ and $\ff(y)$ are shown in \figref{case3}.
\begin{figure}[t]
\centering
\includegraphics[width=0.7\textwidth]{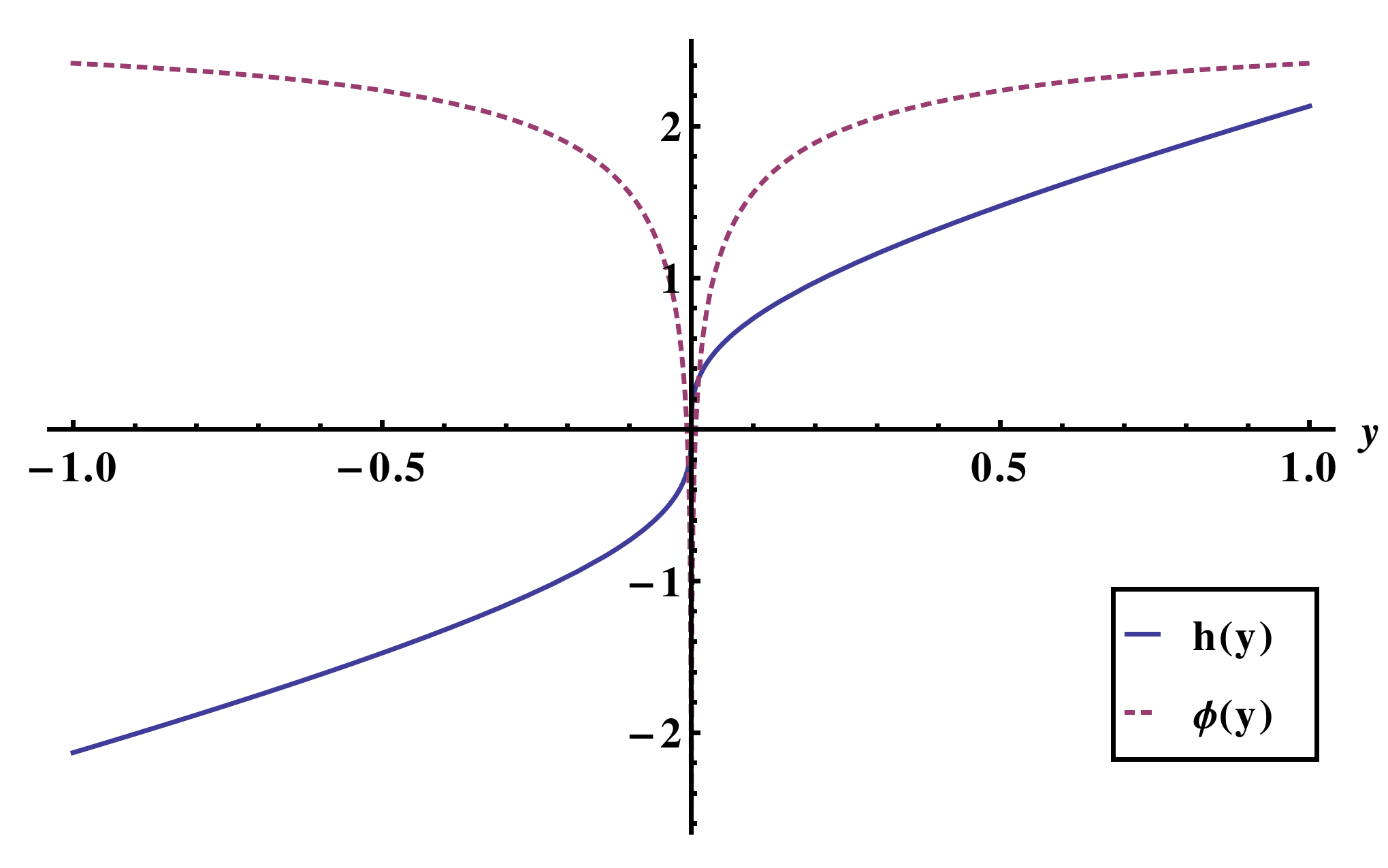}
\caption{Numerical solution in terms of the original functions $h(y)$ and $\ff(y)$ for case 3 ($\y_1 = 0$, $\y_2 = 1$). We have adjusted the free constants to $\a'=4$, $y_0=0$ and $\ff_0 = 4$ for concreteness. The solution becomes singular at the point $y=y_0=0$, which is the physical location of the domain wall.}
\figlabel{case3}
\end{figure}
We remark that this case can be obtained from case~2 by applying the transformation $\a' \to -\a'$ (and using that $\artanh(\im z) = \im\arctan(z)$). This is due to the fact that cases~2 and~3 are related by interchanging $\y_1$ and $\y_2$, which amounts to a sign flip of the order $\a'$ term in~\eqref*{solf}. In the same way, case~5 is related to case~4, and case~7 is related to case~6, as can be seen below.

\paragraph{Case 4: $\y_1 = \text{kink}$, $\y_2 = 0$.}
The solution in this case becomes
\begin{equation}
 \eop^{2f} = \eop^{2(\tau-\tau_0)} + \sfrac{\a'}{16} \left[1 - \tanh(\tau{-}\tau_1)\right]^2 \; , \qquad
 \ff (\tau) = \ff_0 + 2 (f{-}\tau)  \; .
\end{equation}
It is a special case of~\cite{Gemmer:2012pp}, Section 5.2 (with $Q_e = 0$, $a=1$ and taking the decompactification limit in the $S^1$~direction). As $\tau\to +\infty$, the function $f(\tau)$ approaches the linear solution $f(\tau) = \tau$. For $\tau\to -\infty$, the function $f(\tau)$ converges to the constant value $\fr12 \log(\frac{\a'}{4})$. The graph for finite values of $\tau$ qualitatively depends on whether $\tau_0 < \tau_1$ or vice versa. This can be seen from \figref{case4}, where $f(\tau)$ and $\ff(\tau)$ are plotted for different values of the integration constants.
\begin{figure}[t]
\centering
\includegraphics[width=\textwidth]{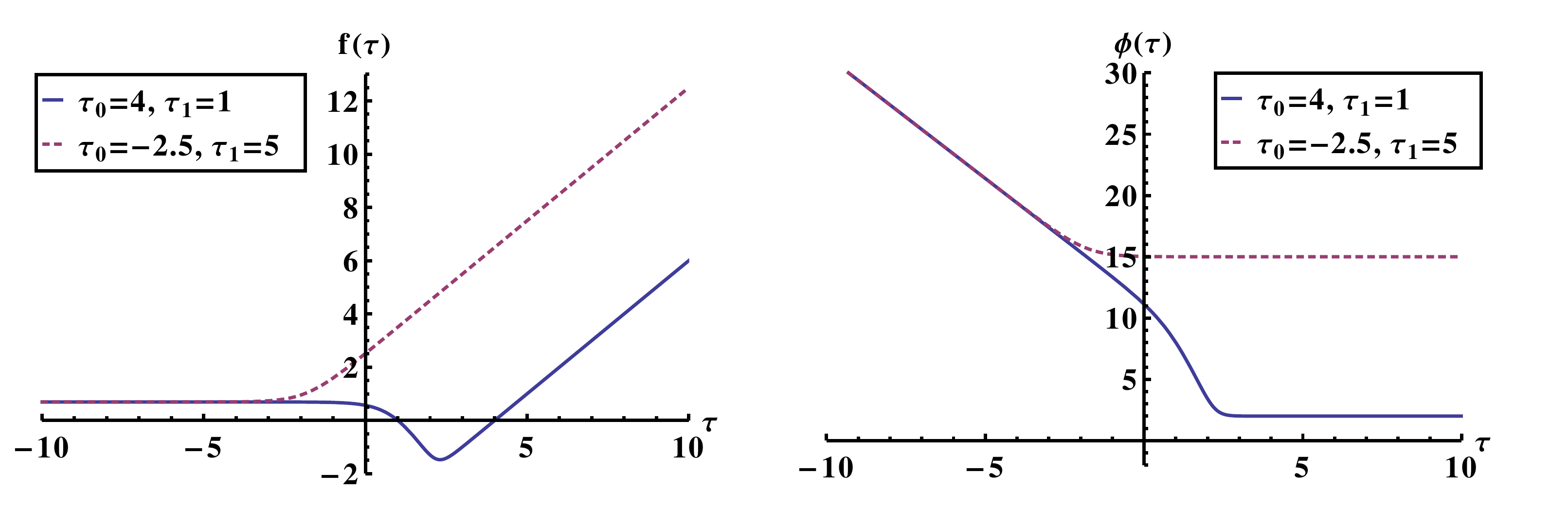}
\caption{Exemplary solutions for case 4 ($\y_1 = \text{kink}$, $\y_2 = 0$) with $\a'=16$ and $\ff_0 = 10$.}
\figlabel{case4}
\end{figure}
We note also that the scalar curvature~\eqref*{scalRtau} remains finite for all $\tau\in\RR$.

\paragraph{Case 5: $\y_1 = 0$, $\y_2 = \text{kink}$.}
The general solution~\eqref*{solf} becomes
\begin{equation}\eqlabel{solf_case5}
 \eop^{2f} = \eop^{2(\tau-\tau_0)} - \sfrac{\a'}{16} \left[1 - \tanh(\tau{-}\tau_2)\right]^2 \; , \qquad
 \ff (\tau) = \ff_0 + 2 (f{-}\tau)  \; .
\end{equation}
The right-hand side approaches zero as $\tau\to\tau_{\text{dw}}$ from above and then becomes negative for sufficiently small values of $\tau$. Hence, $f(\tau)$ derived from expression~\eqref*{solf_case5} is ill-defined for $\tau<\tau_{\text{dw}}$. The limiting value $\tau_{\text{dw}}$ is given by
\begin{equation}\eqlabel{taudw_case5}
 \eop^{\tau_{\text{dw}}} = \frac{\eop^{\fr23(\tau_0 + 2 \tau_2)} X^{\fr13} - 2\,\eop^{2 \tau_2}}{\sqrt{6}\,\eop^{\fr13(\tau_0 + 2 \tau_2)} X^{\fr16}} 
 \; , \quad\text{where}\quad X = 27 \a' + 8\,\eop^{2 (\tau_2-\tau_0)} + \sqrt{27\a' \left(27 \a'{+}16\,\eop^{2 (\tau_2-\tau_0)}\right)} \; .
\end{equation}
In addition, the scalar curvature~\eqref*{scalRtau} diverges as $\tau\to\tau_{\text{dw}}$ from above. The limiting value $\tau_{\text{dw}}$ corresponds to the physical location of the domain wall and our solution exists on the half-space $\tau\in(\tau_{\text{dw}},\infty)$. The graphs of $f(\tau)$ and $\ff(\tau)$ are displayed in \figref{case5}.
\begin{figure}[t]
\centering
\includegraphics[width=0.7\textwidth]{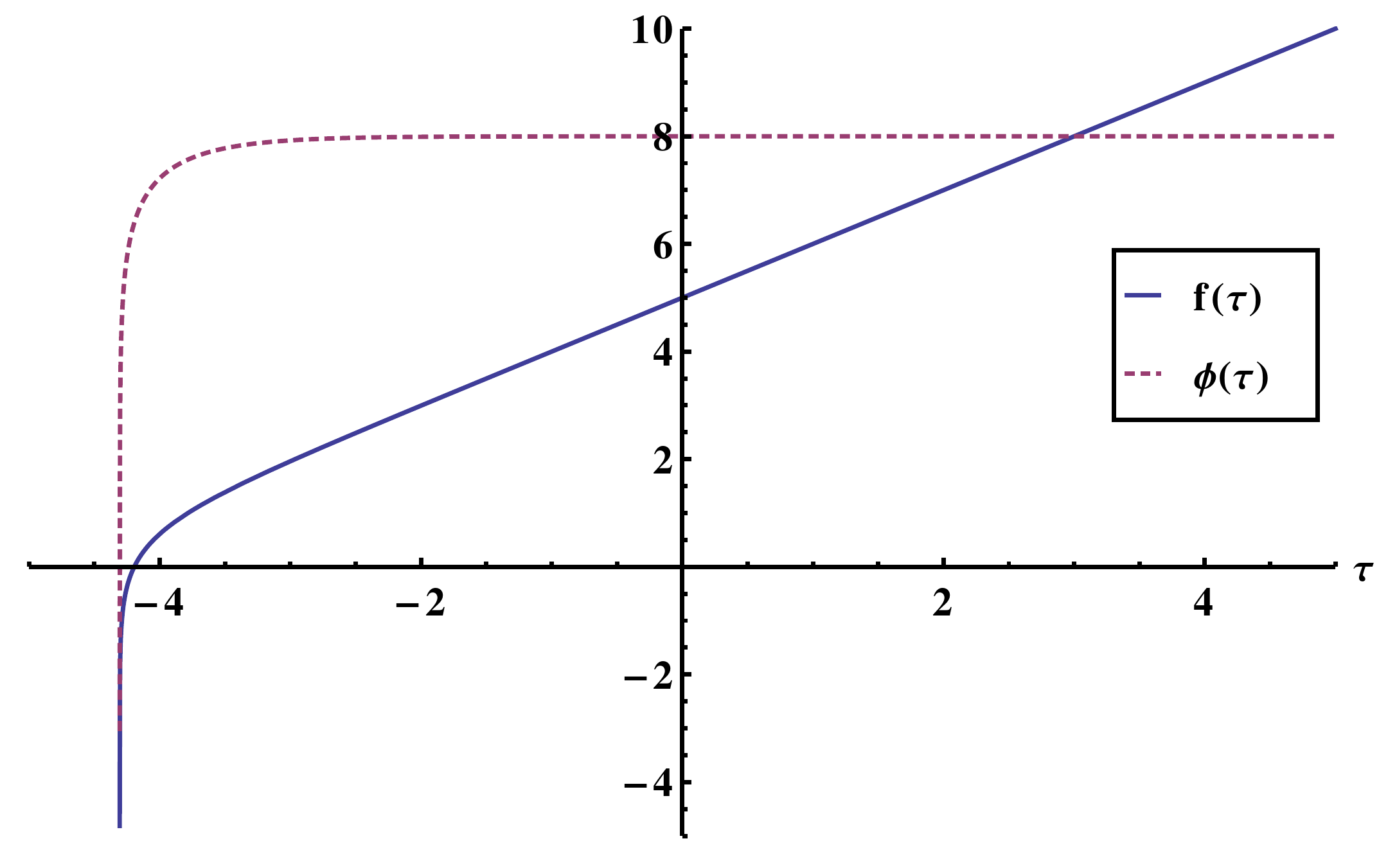}
\caption{Exemplary solution for case 5 ($\y_1 = 0$, $\y_2 = \text{kink}$) with $\a'=16$, $\tau_0 = -5$, $\tau_2 = 3$ and $\ff_0 = -2$.}
\figlabel{case5}
\end{figure}

\paragraph{Case 6: $\y_1 = \text{kink}$, $\y_2 = 1$.}
For this set-up, we obtain from~\eqref*{solf}
\begin{equation}\eqlabel{solf_case6}
 \eop^{2f} = \eop^{2(\tau-\tau_0)} + \sfrac{\a'}{16} \left[\tanh(\tau{-}\tau_1)+1\right] \left[\tanh(\tau{-}\tau_1)-3\right] \; .
\end{equation}
For $\tau\to+\infty$, this solution behaves approximately linear $f(\tau)\approx\tau$. However, depending on the choice of the free parameters, the function $f(\tau)$ may not be well-defined everywhere, because there can be regions where the right-hand side of~\eqref*{solf_case6} becomes negative. We have plotted two qualitatively different scenarios in \figref{case6}. In the plot on the left, the function $f(\tau)$ is well-defined everywhere. In addition, the scalar curvature~\eqref*{scalRtau} has a kink shape implying in particular that it is finite and smooth for all $\tau\in\RR$. The plot on the right is very similar to case 5. In this case, the function $f(\tau)$ derived from~\eqref*{solf_case6} is ill-defined for $\tau<\tau_{\text{dw}}$, where
\begin{equation}\eqlabel{taudw_case6}
 \eop^{\tau_{\text{dw}}} = \frac{\sqrt{\a'\,\eop^{2 \tau_0} + \eop^{\tau_0 + 2\tau_1}\,\sqrt{\a'\,\eop^{\tau_0 - 3\tau_1}}\,\sqrt{\a'\,\eop^{\tau_0 - \tau_1} + 16\,\eop^{\tau_1 - \tau_0}} - 8\,\eop^{2\tau_1}}}{2 \sqrt{2}} \; ,
\end{equation}
and the scalar curvature~\eqref*{scalRtau} diverges as $\tau\to\tau_{\text{dw}}$ from above. As in case 5, the limiting value $\tau_{\text{dw}}$ corresponds to the physical location of the domain wall and our solution exists on the half-space $\tau\in(\tau_{\text{dw}},\infty)$. The distinction between the two scenarios can be made by means of the radicand in the numerator on the right-hand side of expression~\eqref*{taudw_case6}. If it is positive, the solution behaves as shown in the plot on the right. This is the case when $\a' > 2\,\eop^{2(\tau_1 - \tau_0)}$. For $\a' \leq 2\,\eop^{2(\tau_1 - \tau_0)}$ on the other hand, the solution is globally well-defined as shown in the plot on the left. 
\begin{figure}[t]
\centering
\includegraphics[width=\textwidth]{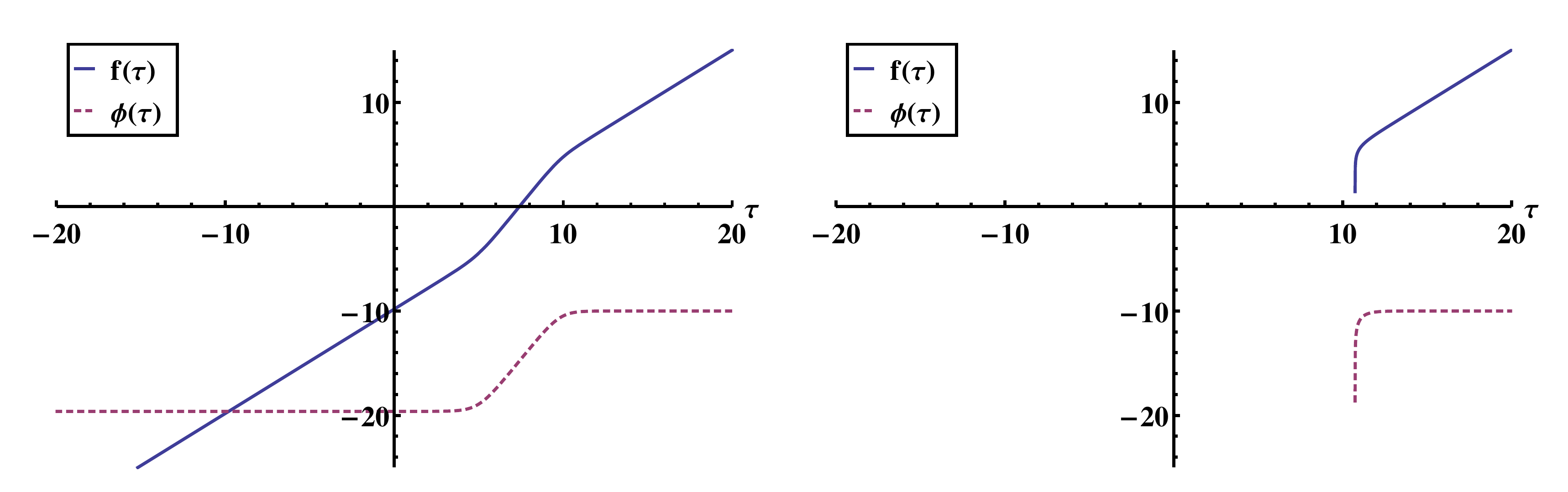}
\caption{Plots of the solution for case 6 ($\y_1 = \text{kink}$, $\y_2 = 1$) for $\a' = 44050$ on the left and $\a' = 4\cdot 10^5$ on the right. The other parameters have been adjusted to $\tau_0 = 5$, $\tau_1 = 10$ and $\ff_0 = 0$. The two graphs have the same asymptotics as $\tau\to+\infty$, but behave qualitatively differently elsewhere.}
\figlabel{case6}
\end{figure}

\paragraph{Case 7: $\y_1 = 1$, $\y_2 = \text{kink}$.}
The general solution~\eqref*{solf} now reads as follows, 
\begin{equation}\eqlabel{solf_case7}
 \eop^{2f} = \eop^{2(\tau-\tau_0)} - \sfrac{\a'}{16} \left[\tanh(\tau{-}\tau_2)+1\right] \left[\tanh(\tau{-}\tau_2)-3\right] \; .
\end{equation}
This case is similar to case 6, except for the flipped sign in front of the order $\a'$ term on the right-hand side. It has already appeared in~\cite{Harland:2011zs} and in~\cite{Gemmer:2012pp}, Section 5.7 as a special case (with $Q_e = 0$, $a=1$ and taking the decompactification limit in the $S^1$~direction). The kink solution~\eqref*{inst_sol} for $\y_2$ approaches the value one, as $\tau\to -\infty$. In this region, the $\a'$ corrections cancel and we recover the zeroth order behavior, that is linear $f$ and constant dilaton. In the limit $\tau\to +\infty$, the expression $\left(\y_1^2 - \y_2^2\right)$ becomes $1$. Thus, for $\tau\gg 1$, the functions $f$ and $\ff$ also become linear and constant, respectively. However, they are shifted by an offset compared to the $\tau\to -\infty$ asymptotics. The $\a'$ corrections are non-constant in an intermediate region and have the important effect of gluing together the different asymptotic functions. This can be seen from \figref{case7}, where the graphs of $f(\tau)$ and $\ff(\tau)$ are plotted for this case. The scalar curvature~\eqref*{scalRtau} is finite and smooth for all $\tau\in\RR$.
\begin{figure}[t]
\centering
\includegraphics[width=0.7\textwidth]{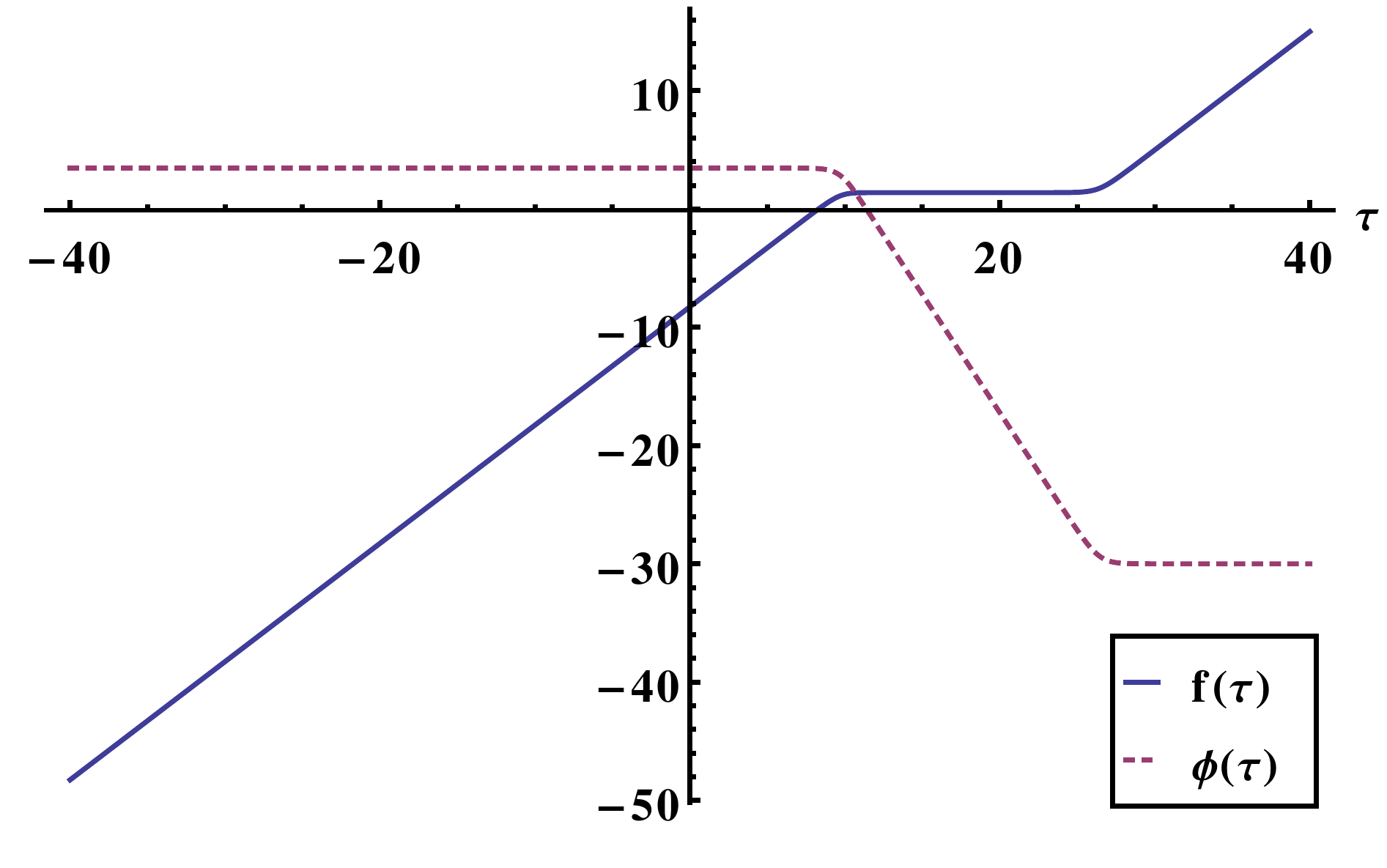}
\caption{Exemplary solution for case 7 ($\y_1 = 1$, $\y_2 = \text{kink}$) with $\a'=64$, $\tau_0 = 25$, $\tau_2 = 10$ and $\ff_0 = 20$.}
\figlabel{case7}
\end{figure}

\paragraph{Case 8: $\y_1 = \text{kink}$, $\y_2 = \text{kink}$, with $\tau_1 \neq \tau_2$.}
The general solution~\eqref*{solf} turns into the following expression,
\begin{equation}\eqlabel{solf_case8}
 \eop^{2f} = \eop^{2(\tau-\tau_0)} + \sfrac{\a'}{16} \left[\tanh^2(\tau{-}\tau_1) - 2 \tanh(\tau{-}\tau_1) - \tanh^2(\tau{-}\tau_2) + 2 \tanh(\tau{-}\tau_2) \right] \; .
\end{equation}
For $\tau\to+\infty$, the contributions from the $\a'$ corrections vanish and we recover the zeroth order behavior, that is linear $f$ and constant dilaton. For $\tau$ finite, the corrections become important and the precise behavior depends on the choice of free parameters. Two qualitatively different scenarios are depicted in \figref{case8}. The plot on the left is an example from a region in parameter space, where the solution resembles that of case 7, with an additional bump, however. In this case the solution is globally well-defined and the scalar curvature~\eqref*{scalRtau} is finite and smooth everywhere. In other regions of parameter space, the solution is similar to that of case 5. This can be seen from the plot on the right. Here, the function $f(\tau)$ is ill-defined for $\tau<\tau_{\text{dw}}$, where $\tau_{\text{dw}}$ is the log of the largest real root of the octic equation
\begin{equation}\eqlabel{taudw_case8}
 4 \left(b^2+x^2\right)^2 \left(c^2+x^2\right)^2 + \a' a^2 (b^2 - c^2) \left( (b^2 + c^2) x^2 + 2 b^2 c^2\right) = 0\; ,
\end{equation}
with $x=\eop^{\tau_{\text{dw}}}$, $a=\eop^{\tau_0}$, $b=\eop^{\tau_1}$, $c=\eop^{\tau_2}$. The closed-form expression for $\tau_{\text{dw}}$ is very lengthy and is thus omitted here for the sake of brevity. The limiting value $\tau_{\text{dw}}$ corresponds to the physical location of the domain wall and the scalar curvature~\eqref*{scalRtau} diverges as $\tau\to\tau_{\text{dw}}$ from above. Our solution exists on the half-space $\tau\in(\tau_{\text{dw}},\infty)$. On the other hand, if there is no solution of~\eqref*{taudw_case8} over the positive reals, we are in a region of parameter space where a globally well-defined solution, such as the one shown in the plot on the left, exists.
\begin{figure}[t]
\centering
\includegraphics[width=\textwidth]{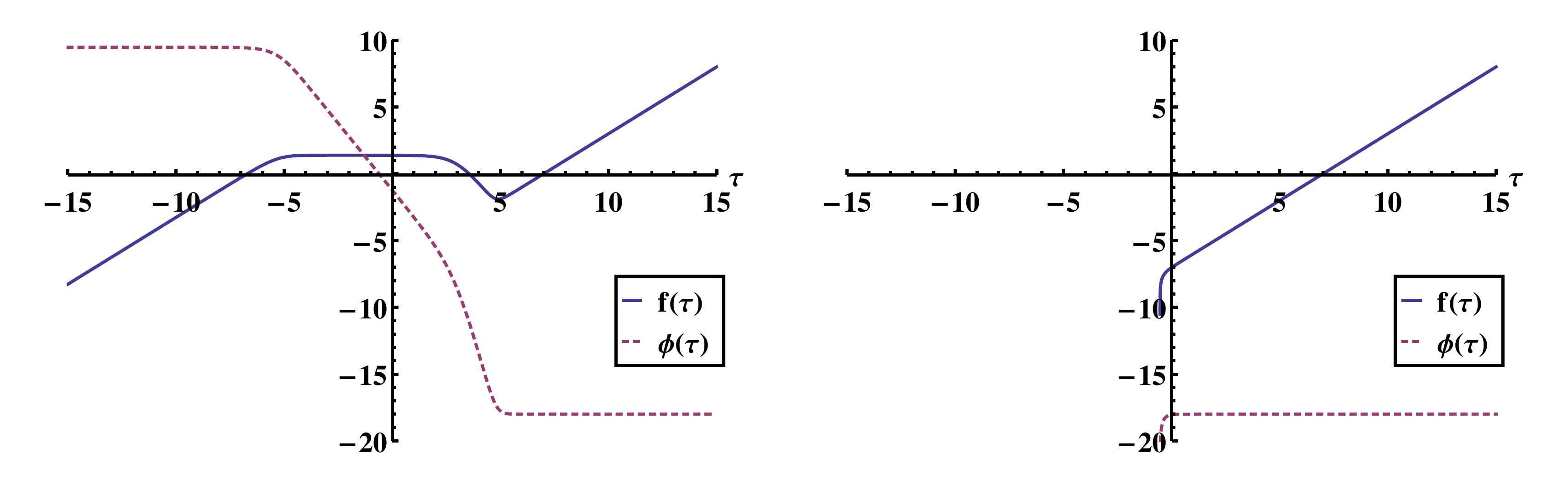}
\caption{Plots of the solution for case 8 ($\y_1 = \text{kink}$, $\y_2 = \text{kink}$, with $\tau_1 \neq \tau_2$) for $\tau_1 = 3$ on the left and $\tau_1 = -6$ on the right. The other parameters have been adjusted to $\a' = 64$, $\tau_0 = 7$, $\tau_2 = -5$ and $\ff_0 = -4$. The two graphs have the same asymptotics as $\tau\to+\infty$, but behave qualitatively differently elsewhere.}
\figlabel{case8}
\end{figure}

\section{Conclusions and outlook}\seclabel{concl}

In this paper, we have studied $(1{+}3)$-dimensional domain wall solutions of heterotic supergravity on a six-dimensional warped nearly K\"ahler manifold $X_6$ in the presence of gravitational and gauge instantons of the type constructed in~\cite{Harland:2011zs}. The instanton contributions are necessary for solving the Yang-Mills sector and the Bianchi identity~\eqref*{BI} at order $\a'$, which is the order we have considered. The ten-dimensional solutions constructed in this paper are of the form
\begin{align}
 \hat{g} &= \h_{\a\b} \dd x^\a \dd x^\b + \dd y^2 + (h(y))^2 \tilde{g}_{uv}(x^w) \dd x^u \dd x^v \; , \\
 \hat{H} &= H + \dd y \wedge H_y \; , \\
 \hat{\ff} &= \ff(y) \; .
\end{align}
where all fields only depend on the non-compact coordinate $y$ transverse to the domain wall. Our solutions preserve two real supercharges, which corresponds to $\mathcal{N}{=}1/2$ supersymmetry from the viewpoint of the four non-compact dimensions spanned by $\{x^\alpha, y\}$.

Following the general formalism developed in~\cite{Lukas:2010mf,Gray:2012md}, we introduced a pair of $y$-dependent $SU(3)$ structure forms $(J, \W)$ on $X_6$ defined via the parallel spinors $\h_\pm$ and rewrote the BPS equations~\eqref*{BPS} as a set of compatibility relations involving $J$, $\W$, $H$, $H_y$ and $\ff$. There is also a static, that is $y$-independent, $SU(3)$ structure denoted $(\tilde{J}, \tilde{\W})$ on $X_6$. The two structures $(J, \W)$ and $(\tilde{J}, \tilde{\W})$ are related by means of the warp factor $h(y)$ and a $y$-dependent mixing angle $\b$.

The BPS equations and Bianchi identity then reduce to a set of ordinary differential equations involving the free functions $h$, $\b$ and $\ff$. The complete system of coupled non-linear ordinary differential equations, as summarized in~\eqrangeref*{fulleqsys1}{fulleqsys6}, is too complicated to solve in full generality. Instead, we studied the special branches $W_1^+ = 0$ and $W_1^- = 0$. While the case $W_1^+ = 0$ can be quickly discarded for it leads to a singular metric, the second case, $W_1^- = 0$, allows for the construction of a variety of solutions depending on the choice of instantons $\y_{1,2}$ for the gravitational and gauge sector. In total, there are eight distinct cases, including already known solutions (cases 1-4 and 7) from~\cite{Harland:2011zs,Gemmer:2012pp,Klaput:2012vv} as well as some new ones (cases 5, 6 and 8).

For the solutions with $W_1^- = 0$, the ten-dimensional fields take the following simpler form,
\begin{align}
 \hat{g} &= \h_{\a\b} \dd x^\a \dd x^\b + \eop^{2f(\tau)} \left( \dd\tau^2 + \tilde{g}_{uv}(x^w) \dd x^u \dd x^v \right) \; , \\
 \hat{H} &= - \sfrac{\a'}{4} \left(\y_1^2 (2 \y_1 - 3)- \y_2^2 (2 \y_2 - 3)\right) \tilde{\W}_+ \; , \\
 \ff &= \ff_0 + 2 (f - \tau) \; , \\
 \eop^{2f} &= \eop^{2(\tau-\tau_0)} + \sfrac{\a'}{4} \left( \y_1^2 - \y_2^2 \right) \; ,
\end{align}
where $\y_{1,2}$ are either $0$, $1$ or the kink solution~\eqref*{inst_sol}, and we used the convenient reparameterization $\dd y = \eop^{f(\tau)} \dd\tau$, $h(y) = \eop^{f(\tau(y))}$. In case one (equivalent to zeroth order in $\alpha'$), $f = \tau{-}\tau_0$, which leads to a cone metric with constant dilaton and vanishing NS 3-form flux. The other cases typically asymptote to this zeroth order behavior either at $\tau\to+\infty$, $\tau\to-\infty$ or $\tau\to\pm\infty$. Close to the domain wall, care must be taken as to the validity of the supergravity approximation. Indeed, in some of the cases, the scalar curvature diverges as the domain wall is approached.

The NS 3-form flux $\hat{H}$ is always proportional to $\tilde{\W}_+$ in all our explicit constructions. It would be interesting to have access to solutions with a more general $\hat{H}$ that also includes terms proportional to $\tilde{\W}_-$ and $\tilde{J}\wedge \dd y$. It remains to be seen whether this can be achieved by finding a solution of the general system~\eqrangeref*{fulleqsys1}{fulleqsys6} with both $W_1^+ \neq 0$ and $W_1^- \neq 0$ or whether another, perhaps rather different, ansatz is necessary. This is left for future work.

\section*{Acknowledgments}

We would like to thank Alexander~Popov for useful discussions and for
collaboration  in the early stages of this work. We are grateful to the referee
for careful reading and constructive comments. This work was partially supported
by the Deutsche Forschungsgemeinschaft grant LE 838/13. A.S.H. thanks the
Albert-Einstein-Institute and its director Hermann~Nicolai for warm hospitality
and generous financial support through the award of a Max Planck Society
research stipend.  The work of E.T.M. was supported by the Riemann Center for
Geometry and Physics, \'Ecole Normale Sup\'erieure de Lyon and by the
Albert-Einstein-Institute. E.T.M. personally thanks Henning Samtleben and
Hermann Nicolai for warm hospitality during completion of part of this project.

\section*{Appendix}
\appendix
\section{Conventions}\applabel{conv}

In this paper we use the following conventions for indices and normalizations. The full range of $(1{+}9)$-dimensional indices is split by the presence of the domain wall and will be distinguished by means of the following set of Greek and Latin letters,
\begin{equation}
\begin{aligned}
&\m,\n,\r,\s && =0,1,\ldots,9 \; ,\\
&\a,\b,\g,\d,\e&& =0,1,2 \; ,\\
&a,b,c,\ldots,m,n && = 3,4,\ldots,9 \; ,\\
&u,v,w,x,y,z && = 4,5,\ldots,9 \; .
\end{aligned}
\end{equation}
These are understood as curved space-time indices. In addition, we sometimes need to use tangent space (local Lorentz) indices, which are denoted by underlined indices.
  
(Anti-)symmetrization is always performed with a factor of $(1/n!)$, that is with weight one. For example, we define
\begin{equation}
A_{[\m}B_{\n]}:=\sfr{1}{2!}(A_\m B_\n-B_\n A_\m) \; ,
\end{equation}
for the case $n=2$. A $p$-form $\w$ is expanded into components according to
\begin{equation}
 \w := \sfr{1}{p!} \w_{\m_1 \ldots \m_p} \dd x^{\mu_1} \wedge\cdots\wedge \dd x^{\mu_p} \; .
\end{equation}
The Clifford action of a $p$-form $\omega$ on a spinor $\epsilon$ is defined as 
\begin{equation}
 \omega \cdot \epsilon := \sfrac 1{p!} \omega_{\mu_1 \dots \mu_p} \gamma^{\mu_1 \dots \mu_p} \epsilon,
\end{equation}
where $\gamma^{\mu_1 \dots \mu_p} := \gamma^{\left[\mu_1\right.}\cdots\gamma^{\left. \mu_p\right]}$ and we use the Clifford algebra convention $\{\gamma^\mu,\gamma^\nu\} = 2g^{\mu\nu}$ for the higher-dimensional gamma matrices $\g^\mu$.

A connection $\nabla$ on a manifold ${\cal M}$ with vielbein $\s^{\ba}_b$ is defined to act on vectors $v^b$ and spinors $\e$ in the following way,
\begin{equation}\eqlabel{defconn}
\begin{aligned}
\nabla_a v^b&=\dt_av^b + \G_{a}{}^b{}_{c}v^c \; ,\\
\nabla_a \e&=\dt_a \e + \sfr14\w_a{}\cdot \e \; , \qquad \w_a=\sfr12 \w_a{}^{\bc}{}_{\bb}\g_{\bc}{}^{\bb} \; .
\end{aligned}
\end{equation}
The components of the spin connection $\w_a{}^{\bb}{}_{\bc}$ are related to $\G_{a}{}^b{}_{c}$ via
\begin{equation}\eqlabel{defspinconn}
\w_a{}^{\bb}{}_{\bc}=\G_{a}{}^{\bb}{}_{\bc}-\s^d_{\bc}\dt_a \s^{\bb}_d \; .
\end{equation}
The torsion $T^{\ba}$ of the connection $\nabla$ is defined as $T^{\ba}=\nabla \s^{\ba}$. Using~\eqrangeref*{defconn}{defspinconn}, we can expand the torsion into components as follows,
\begin{equation}
\begin{aligned}
T^{\ba}&=\nabla \s^{\ba} = \dd\s^{\ba}+\w^{\ba}{}_{\bb}\wedge \s^{\ba} = \G_{\bb}{}^{\ba}{}_{\bc} \, \s^{\bb}\wedge \s^{\bc} \; , \qquad \text{or}\\
T_{bc}{}^a&=\G_{[b}{}^a{}_{c]} \; ,
\end{aligned}
\end{equation}
where $\w^{\ba}{}_{\bb}=\w_c{}^{\ba}{}_{\bb} \, \dd x^c$ is the connection 1-form and the $\s^{\ba}=\s^{\ba}_b \, \dd x^b$ define an orthonormal frame on ${\cal M}$.

\bibliographystyle{utphys}
\bibliography{domwall}

\end{document}